\documentclass[a4paper,twocolumn,aps,prb,superscriptaddress,showpacs,floatfix]{revtex4}
\usepackage{amssymb,amsmath,stmaryrd}
\usepackage{times}

\usepackage{array}
\usepackage{graphicx}
\usepackage{ifpdf}
\usepackage{color}

\newcommand{\be}{\begin{equation}}
\newcommand{\ee}{\end{equation}}
\newcommand{\bea}{\begin{eqnarray}}
\newcommand{\eea}{\end{eqnarray}}

\def\a{\alpha}
\def\b{\beta}
\def\g{\gamma}
\def\G{\Gamma}
\def\d{\delta}
\def\D{\Delta}

\def\ve{\varepsilon}

\def\th{\theta}

\def\l{\lambda}

\def\n{\nu}
\def\c{\xi}

\def\p{\pi}

\def\r{\rho}

\def\t{\tau}

\def\F{\Phi}

\def\w{\omega}
\def\W{\Omega}

\def\Q{\Psi}

\def\bgve{\mbox{\boldmath $\varepsilon$}}

\def\blb{{\mathbf b}}

\def\bld{{\mathbf d}}
\def\ble{{\mathbf e}}

\def\blk{{\mathbf k}}

\def\bln{{\mathbf n}}

\def\blr{{\mathbf r}}

\def\blB{{\mathbf B}}

\def\blE{{\mathbf E}}

\def\blJ{{\mathbf J}}
\def\blK{{\mathbf K}}

\def\blN{{\mathbf N}}

\def\callE{\mbox{$\mathcal{E}$}}

\def\callH{\mbox{$\mathcal{H}$}}

\def\callL{\mbox{$\mathcal{L}$}}

\def\callP{\mbox{$\mathcal{P}$}}

\def\callS{\mbox{$\mathcal{S}$}}

\def\callU{\mbox{$\mathcal{U}$}}

\def\bcallE{\mbox{\boldmath $\mathcal{E}$}}

\def\ra{\rightarrow}

\def\de{\partial}
\def\iif{\infty}
\def\bra{\langle}
\def\ket{\rangle}
\def\grad{\mbox{\boldmath $\nabla$}}
\def\Tr{{\rm Tr}}
\def\Re{{\rm Re}}
\def\Im{{\rm Im}}

\def\1op{\hat{\mathbbm{1}}}
\def\nn{\nonumber}

\begin{document}

\title{Some Exact Properties of the Nonequilibrium Response Function 
for Transient Photoabsorption}

\author{E. Perfetto}
\affiliation{Dipartimento di Fisica, Universit\`{a} di Roma Tor Vergata,
Via della Ricerca Scientifica 1, 00133 Rome, Italy}
\affiliation{INFN, Laboratori Nazionali di Frascati, Via E. Fermi 40, 00044 Frascati, 
Italy}
\author{G. Stefanucci}
\affiliation{Dipartimento di Fisica, Universit\`{a} di Roma Tor Vergata,
Via della Ricerca Scientifica 1, 00133 Rome, Italy; European Theoretical Spectroscopy Facility (ETSF)}
\affiliation{INFN, Laboratori Nazionali di Frascati, Via E. Fermi 40, 00044 Frascati, 
Italy}

\begin{abstract}
The physical interpretation of time-resolved photoabsorption 
experiments is not as straightforward as for the more conventional photoabsorption 
experiments conducted on equilibrium systems. In fact, the relation 
between the transient photoabsorption spectrum and the properties of the examined sample  
can be rather intricate 
since the former is a complicated functional of both the driving pump and the feeble probe 
fields. In this work we critically review the derivation of the 
time-resolved photoabsorption spectrum in terms of the nonequilibrium  
dipole response function $\chi$ and assess its domain of validity. We 
then analyze $\chi$ in detail and discuss a few exact
properties useful to interpret the transient spectrum {\em during} 
(overlapping regime)
and {\em after} (nonoverlapping regime) the action of the pump. 
The nonoverlapping regime is the simplest to address. The absorption 
energies are indeed independent of the delay between the pump and probe 
pulses and hence the transient spectrum can change only by a 
rearrangement of the spectral weights. We give a close expression of 
these spectral weights in two limiting 
cases (ultrashort and everlasting monochromatic probes) and highlight 
their strong dependence on coherence and probe-envelope.  
  In the overlapping regime we obtain a Lehmann-like representation of 
$\chi$ in terms of light-dressed states and provide a unifying framework 
of various well known effects in pump-driven systems. We also 
show the emergence of spectral sub-structures due to the 
finite duration of the pump pulse.

\end{abstract}


\maketitle

\section{Introduction}

Filming a ``movie'' with electrons and 
nuclei as actors may sound fantasy-like, but  it 
is {\em de facto} a common practice in physics and chemistry 
modern laboratories. With the impressive march of advances in laser 
technology, 
ultrashort (down to the sub-fs time-scale), intense ($\gtrsim 
10^{15}$ W/cm$^{2}$) and focussed pulses 
of designable shape, hereafter named {\em pumps}, are available 
to move electrons in real or energy space. By recording the 
photoemission or photoabsorption spectrum produced by a second, weak 
pulse, hereafter named {\em probe}, impacting the sample with a 
tunable delay from the pump a large 
variety of ultrafast physical and chemical 
processes can be documented.\cite{ki.2009,bgk.2009,gck.2012,spsk.2012,ghlsllk.2013,kc.2014}
Time-resolved pump-and-probe (P\&P) spectroscopies have revealed 
the formation and dynamics of 
excitons,\cite{kchlc.2003,wdbh.2005,kkkg.2006,pdbletal.2009,cblh.2012} 
charge-transfer 
excitations,\cite{may-book,ahwl.2000,sbpotall.2002,ffhfmetal.2005,gw.2005,rl.2006,rfsrmetal.2013,frbmetal.2014,pdrfclr.2015} 
auto-ionized states\cite{pcsdk.2011,bblwbetal.2014} and light-dressed 
states,\cite{bsy.2007,ghsaetal.2010,gcfkwc.2010,rthzetal.2011,tg.2012,lpnlg.2012,cbbmetal.2012,czwcetal.2012,cwgs.2013,cwgs2.2013,hwlsetal.2013,cwcwetal.2013,pbbmnl.2013,wccwtc.2013,bbwwetal.2014} 
the evolution of Fano 
resonances,\cite{wbkd.2005,zl.2012,cl.2013,okrmetal.2013,okaretal.2014,aopm.2014} the screening build-up of 
charged excitations,\cite{htbbal.2001,hkcp.2003,bsme.2004,mpb.2012} 
the transient transparency of solids,\cite{sswk.2004,netal.2009} the motion of valence 
electrons,\cite{dhkuetal.2002,nvc.2005,smpdetal.2009,spmdi.2009,glwsetal.2010,pghmetal.2011} the band-gap 
renormalization of excited 
semiconductors,\cite{bre.2009,sbshetal.2013,srpswetal.2014} how 
chemical bonds break\cite{z.2000,c.2005,skpmetal.2010,ls.2003}
and other fundamental phenomena. 

A suited P\&P spectroscopy to investigate charge-neutral excitations is 
the time-resolved (TR) photoabsorbtion (PA) 
spectroscopy.\cite{cmuetal.2007,wcczhetal.2010,hsgk.2011} It is well 
established that PA spectra of equilibrium systems are proportional 
to the dipole-dipole response function 
$\chi$,\cite{bassani-book,s.1988,vignale-book,svl-book} an extremely useful quantity 
to understand and interpret the experimental results.
In pump-driven systems the derivation 
of a mathematical quantity to interpret TR-PA spectra is slightly more 
delicate and, in fact, several recent works 
have been devoted to this subject.\cite{gbts.2011,sypl.2011,cl.2012,blm.2012,dgc.2013}
The difficulty in constructing a solid and general TR-PA spectroscopy 
framework (valid for general P\&P envelops, durations and delays and 
for samples of any thickness)
stems from the fact that the probed systems evolve in 
a strong {\em time-dependent} electromagnetic (em) field  and hence 
(i) low-order 
perturbation theory in the pump intensity may not be sufficiently 
accurate 
and (ii) separating the total energy per unit frequency absorbed by the system into a pump and 
probe contribution is questionable. Furthermore,  due  to the lack 
of time-translational invariance the TR-PA spectrum is not an intrinsic 
property of the pump-driven system, depending it on the shape of 
the probe field too. 

We can distinguish two different approaches to derive a 
TR-PA formula: the energy approach,\cite{gbts.2011,cl.2012,dgc.2013} which aims at 
calculating the energy absorbed from only the probe, and 
the Maxwell approach,\cite{sypl.2011,blm.2012,mukamel-book} which aims at calculating the transmitted probe 
field (these approaches  are equivalent for optically thin and
equilibrium samples). We carefully revisit the energy approach, 
highlight its limitations and infer that it is not suited to 
perform a spectral decomposition of the absorbed energy.
We also re-examine the Maxwell approach and 
provide a derivation of the TR-PA spectrum in non-magnetic 
systems without the need of frequently made assumptions like, e.g., 
slowly-varying probe envelops or ultrathin samples. The final result 
is that the TR-PA spectrum can be calculated from the single and 
double convolution of the nonequilibrium response function $\chi$
with the probe field. 

For the physical interpretation of TR-PA spectral features  
a Lehmann-like representation of the nonequilibrium $\chi$ would be 
highly valuable, as it is in PA spectroscopy of equilibrium systems.
In this work we discuss some exact properties of the nonequilibrium 
$\chi$ and of its convolution with the probe field.
When the probe acts after the pump (nonoverlapping 
regime) $\chi$ can be written as the average over a {\em nonstationary} 
state of the dipole operator-correlator evolving with the 
{\em equilibrium} Hamiltonian of the sample. 
In this regime the TR-PA spectrum is nonvanishing when the frequency 
matches the difference of two excited-state energies. As 
these energies are independent of the delay $\t$ between the pump and 
probe field, only the spectral weights can change with $\t$ (not 
the absorption regions).\cite{psms.20xx} 
We discuss in detail how the spectral weights are 
affected by the coherence between nondegenerate excitations 
and by the shape of the probe field. A close expression is given in the 
two limiting cases of 
ultrashort and everlasting monochromatic probes.

The overlapping regime is, in general, much more complicated to 
address. The absorption energies cease to be an intrinsic property 
of the unperturbed system and acquire a dependence on the delay.
Nevertheless, an analytic treatment is still possible in 
some relevant situations. 
For  many-cycle pump fields of duration longer than the typical dipole 
relaxation time, we show that a Lehmann-like representation of the 
nonequilibrium $\chi$ in terms of light-dressed states can be used to 
interpret the TR-PA spectrum.
We provide a unifying
framework of well known effects in pump-driven systems like, e.g., the AC 
Stark shift, the Autler-Townes 
splitting and the Mollow triplet. 
More analytic results can be found for samples described 
by a few level systems. 
In this case, from the {\em  exact} solution of the nonequilibrium 
response function with 
pump fields of finite duration we obtain the dipole moment induced by 
ultrashort probe fields. The analytic expression shows 
that (i) the $\t$-dependent 
renormalization of the absorption energies 
follows closely the pump envelope and (ii)
a spectral sub-structure characterized by extra absorption energies 
emerges.

The paper is organized as follows. In Section~\ref{TR-PASsec} we 
briefly review the principles of
TR-PA spectroscopy measurements. 
We critically discuss the energy approach to PA spectroscopy in 
equilibrium systems and highlight the limitations which hinder a 
generalization to nonequilibrium situations. The conceptual problems 
of the energy approach are overcome by the Maxwell approach which is 
re-examined and used to derive the transmitted probe field emerging 
from non-magnetic samples of arbitrary thickness, without any 
assumption on the shape of the incident pulse. The nonequilibrium 
dipole-dipole response function $\chi$ is introduced in 
Section~\ref{nechisec} and related to the transmitted probe field. 
We analyze $\chi$ in the nonoverlapping regime in 
Section~\ref{nonoverlapsec} and in the overlapping regime in 
Section~\ref{overlapsec}. Finally, a class of exact solutions in few-level 
systems for overlapping pump and probe fields 
is presented in Section~\ref{numsec}. Summary and 
conclusions are drawn in Section~\ref{summsec}.

\section{Time-Resolved Photoabsorption Spectroscopy}
\label{TR-PASsec}

In this section we briefly revisit the principles of PA
for systems in equilibrium, and subsequently generalize the discussion to the more recent 
TR-PA for systems driven away from equilibrium. 
The aim of this preliminary section is to highlight the underlying 
assumptions of orthodox equilibrium PA theories, identify the 
new physical ingredients that a nonequilibrium PA theory should 
incorporate, and eventually obtain a formula for the TR spectrum 
which is a functional of {\em both} the pump and probe fields.
Except for a critical review of the literature no original 
results are present in this 
section.

\subsection{Experimental measurement}
\label{expsec}

\begin{figure}[tbp]
	\centering
       \includegraphics[width=0.9\linewidth]{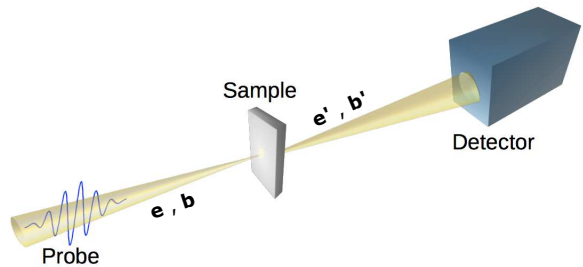}
       \caption{(Color online) Illustration of a PA experiment.}
      \label{eqPA}
\end{figure}

Consider a system in equilibrium and irradiate it with some feeble 
light (perturbative probe). In Fig. \ref{eqPA} 
we show a snapshot at 
time $t$ of a typical equilibrium PA experiment. The incident light 
is described by the electric and magnetic fields $\ble(t)$ and 
$\blb(t)$  (left side of the 
sample) whereas the transmitted light is described by the  
em
fields $\ble'(t)$, $\blb'(t)$  (right side of the sample). 
The experiment measures the 
total transmitted energy $E'$. This quantity is 
given by the energy flow (or equivalently the 
 Poynting vector) integrated over time (the duration of 
the experiment) and surface.
Denoting by $\callS$ the cross 
section  of the incident beam we have (here and in the following integrals with no 
upper and lower limits go from $-\iif$ to $+\iif$)
\be
E'=\callS\frac{c}{4\p}\int dt \left|
\ble'(t)\times \blb'(t)\right|.
\label{tote}
\ee
The integral in Eq.~(\ref{tote}) is finite since the em fields used in an experiment 
vanish outside a certain time interval. 
In vacuum the electric and magnetic fields are 
perpendicular to each other and their cross product is parallel to the direction of 
propagation. Taking into account that $|\blb'|=|\ble'|$ the 
transmitted energy in Eq. (\ref{tote}) simplifies to
\be
E'=\callS\frac{c}{4\p}\int dt \left|
\ble'(t)\right|^{2}.
\label{tote2}
\ee

From Eq. (\ref{tote2}) we see that the transmitted energy $E'$ depends on the temporal shape of the 
electric field. This dependence can be exploited to extract the energy 
of the neutral excitations of the sample. A typical, systematic way of varying the 
temporal shape consists in probing the sample with monochromatic 
light of varying frequency. Taking into account that the electric field is real, its Fourier 
transform reads
\be
\ble'(t)=\int_{0}^{\iif}\frac{d\w}{2\p}\,
\tilde{\ble}'(\w)e^{-i\w t}+c.c.
\label{eft}
\ee
where $c.c$ stands for ``complex conjugate''.
Unless otherwise defined, quantities with the tilde symbol 
on top denote the Fourier 
transform of the corresponding time-dependent quantities.
Inserting Eq. (\ref{eft}) back into Eq. (\ref{tote2}) we find
\be
E'=\callS\frac{c}{2\p}\int_{0}^{\iif} \frac{d\w}{2\p} \left|
\tilde{\ble}'(\w)\right|^{2}.
\label{transw}
\ee
For monochromatic light of frequency $\w_{0}$ (in the time interval 
of the experiment) the transmitted field  
$\tilde{\ble}'(\w)$ is peaked at $\w\simeq \w_{0}$ and hence 
$
E'\simeq \callS\frac{c}{2\p}\frac{\D\w}{2\p} \left|
\tilde{\ble}'(\w_{0})\right|^{2}
$,
where $\D\w$ is the width of the peaked function $\tilde{\ble}'(\w)$.
Therefore the quantity
\be
\tilde{W}'(\w)\equiv \callS \frac{c}{2\p}\left|
\tilde{\ble}'(\w)\right|^{2}
\quad,\quad\w>0
\ee
can be interpreted as the transmitted energy per unit frequency. 

Alternatively $\tilde{W}'(\w)$ could be measured using fields of arbitrary temporal 
shape and a spectrometer. As this is also the technique in TR-PA 
experiments, and the method to elaborate the data of the real-time 
simulations of Section \ref{numsec} is based on this technique,
we shortly describe its principles. The 
spectrometer, placed between the sample and the detector, splits  
the transmitted beam into two halves and generates a tunable delay 
$\d$ for one of the halves.
The resulting electric field at the detector is 
therefore 
$\frac{1}{2}\left(\ble'(t)+\ble'(t-\d)\right)$, and
the measured transmitted energy  is 
\be
E'(\d)=\callS\frac{c}{4\p}\int dt \left|
\frac{\ble'(t)+\ble'(t-\d)}{2}\right|^{2}.
\ee
The PA experiment is  
repeated for different delays $\d$, and the results are collected  to 
perform a cosine transform 
\be
\tilde{E}'(\n) \equiv \int d\d\; E'(\d)\cos(\n\d).
\label{costra}
\ee
The relation between 
$\tilde{E}'$ and $\tilde{W}'$ is readily found. Using Eq. (\ref{eft}) we 
get
\be
E'(\d)=\callS\frac{c}{2\p}\int_{0}^{\iif} \frac{d\w}{2\p} \left|
\tilde{\ble}'(\w)\right|^{2}\frac{1+\cos(\w\d)}{2}.
\ee
Inserting this result into Eq. (\ref{costra}) and taking into account the 
identity
$\int d\d \cos(\w\d)\cos(\n\d)=\p\left[\d(\w+\n)+\d(\w-\n)\right]$
we find
$
\tilde{E}'(\n)=\p\d(\n)E'+\callS\frac{c}{8\p}\left|\tilde{\ble}'(\n)\right|^{2}.
$
Thus for every $\n\neq 0$ we have
$\tilde{W}'(\n)=4\tilde{E}'(\n)$.

The PA experiment can be repeated {\em without} 
the sample to measure the energy per unit frequency 
$\tilde{W}(\w)$ of the incident beam. The difference
\be
\tilde{S}(\w)=\tilde{W}(\w)-\tilde{W}'(\w)>0
\label{dW}
\ee
is therefore the missing energy per unit frequency. How to relate this 
experimental quantity to the excited energies and excited states of the 
sample is well established and will be reviewed in the next Section.

\begin{figure}[t]
	\centering
       \includegraphics[width=0.9\linewidth]{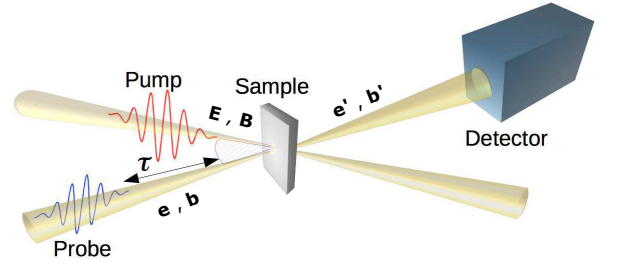}
       \caption{(Color online) Illustration of a TR-PA experiment. }
      \label{nneqPA}
\end{figure}

The main novelty introduced by TR-PA experiments consists in 
probing the sample in a nonstationary (possibly driven) state.
The sample is driven out of equilibrium by an intense laser 
pulse described by the em fields $\blE(t)$ and  
$\blB(t)$, and subsequently probed with the em fields $\ble(t)$ and 
$\blb(t)$, see Fig. \ref{nneqPA}. We refer to $\blE(t)$ and 
$\blB(t)$ as the {\em pump} fields.
To extract information on the missing probe energy 
the transmitted pump field is {\em not} measured, see 
again Fig. \ref{nneqPA}. As the sample is not in its ground state the 
transmitted beam is also made of photons produced by the stimulated 
emission.
These photons have the same frequency and direction of the probe 
photons and, therefore, the inequality in Eq. (\ref{dW}) is no longer 
guaranteed.

\subsection{Energy approach: PA in equilibrium}

In this section we obtain an expression for the spectrum $\tilde{S}(\w)$ of systems 
initially in equilibrium in their ground state (the 
finite-temperature generalization is straightforward). We use an approach 
based on the energy dissipated by the sample and highlight those 
parts in the derivation where the hypothesis of initial equilibrium
and weak em field  is used. We advance that this approach cannot be 
generalized to nonequilibrium situations.

For a system driven out of 
equilibrium by an external (transverse) electric field $\blE_{\rm ext}$ 
the energy absorbed per unit time, i.e., the
power dissipated by the system, is 
\be
\callP(t)=-\int d\blr \;\blJ(\blr t)\cdot \blE_{\rm ext}( \blr t),
\label{power}
\ee
where $\blJ$ is the current density and $e=-1$ is the electric charge.
This well-known formula is valid only provided that the electric field generated by 
the induced current $\blJ$ is much smaller than $\blE_{\rm ext}$. For 
the time being let us assume that this is the case.
We also assume that the probed systems are   
nanoscale samples 
like atoms and molecules or thin slabs of solids. Then the wavelength of 
the incident em field is typically much 
larger than the longitudinal dimension of the sample and 
the spatial dependence of $\blE_{\rm ext}$ can be ignored.
Writing 
$\blJ=\grad(\blr\cdot\blJ)-(\grad\cdot\blJ)\blr$, 
discarding the total divergence and using the continuity equation the power becomes 
\be
\callP(t)
=-\int d\blr \;\frac{\de n(\blr 
t)}{\de t}\blr\cdot \blE_{\rm ext}( t),
\label{manp}
\ee
where $n$ is the electron density. The integral of the power between any 
two times $t_{1}$ and $t_{2}$ yields the difference between the 
energy of the system at time $t_{2}$ and the energy of the system at 
time $t_{1}$:
\be
E_{\rm sys}(t_{2})-E_{\rm sys}(t_{1})=
-\int_{t_{1}}^{t_{2}} dt \;\blE_{\rm ext}( t)\cdot \frac{d}{dt}
\bld(t) ,
\label{e2e1}
\ee
where we found convenient to define the dipole moment
\be
\bld(t) =\int d\blr\; \blr \,n(\blr t).
\label{defdipave}
\ee

We observe that nowhere 
the assumption of small external em fields and/or the assumption of a system in 
equilibrium are made in the derivation of Eq. 
(\ref{e2e1}). It is also important to emphasize the semiclassical nature 
of Eq. (\ref{e2e1}). Suppose that the sample is 
initially in its ground state with energy $E_{g}$. We switch the em field on at a time 
$t=t_{\rm on}$ and switch it off at 
a time $t=t_{\rm off}$. According to Eq. (\ref{e2e1}) the difference 
$E_{\rm sys}(t)-E_{g}$ remains constant for any 
$t>t_{\rm off}$.  Physically, however, this is not what happens.
At times $t\gg t_{\rm off}$ the  sample is back in its ground 
state since there has been enough time to relax (via the spontaneous emission of 
light). Hence the correct physical result should be 
$E_{\rm sys}(t\to\iif)-E_{g}=0$. In Eq.~(\ref{e2e1}) the description of the em field is 
purely classical and  
does not capture the phenomenon of spontaneous emission. 
Nevertheless, a spontaneous emission 
process occurs on a time scale much longer 
than the duration of a typical PA experiment. 
The semiclassical formula is therefore accurate for 
times  $t\gtrsim t_{\rm off}$ and, consequently, the quantity
\be
E_{\rm abs}=
-\int dt \;\blE_{\rm ext}( t)\cdot \frac{d}{dt}\bld( t) 
\label{DeltaE}
\ee
can be identified with the increase in the energy of the sample just after the em field 
has been switched off. We refer to $E_{\rm abs}$ as the 
{\em absorbed energy}. Let us see 
how to relate Eq. (\ref{DeltaE}) to
the missing energy $\tilde{S}(\w)$ measured in an experiment. 

In equilibrium PA experiments $\blE_{\rm ext}=\ble$ is the probing 
field discussed in Section \ref{expsec}. 
Let  $E=\int\frac{d\w}{2\p}\tilde{W}(\w)$ and  
$E'=\int\frac{d\w}{2\p}\tilde{W}'(\w)$ be the total energy of the incident and 
transmitted beam  respectively. Then the difference $E-E'$ is 
the energy transferred to the sample, i.e., the absorbed energy of Eq. 
(\ref{DeltaE})
\be
E-E'=E_{\rm abs}.
\label{preldisc}
\ee
We write the dipole moment $\bld=\bld_{\rm eq}+\bld_{p}$ as the the 
sum of the equilibrium value $\bld_{{\rm eq}}$ and the probe induced 
variation $\bld_{p}$. Since $\bld_{{\rm eq}}$ is constant in time the 
absorbed energy 
in frequency space reads 
\be
E_{\rm abs}=i\int_{0}^{\iif} \frac{d\w}{2\p} \w \;
\tilde{\ble}^{\ast}(\w)\cdot\tilde{\bld}_{p}(\w)+c.c .
\label{spasabs}
\ee
Taking into account Eq. (\ref{preldisc}) and the 
definition of $\tilde{S}$ in Eq. (\ref{dW})  we also have
\be
E_{\rm abs}= \int\frac{d\w}{2\p} \tilde{S}(\w).
\label{eabs2}
\ee
We now  show that the r.h.s. of Eqs. (\ref{spasabs}) and (\ref{eabs2}) are the same because the integrands are the same. 
The transmitted em field $\tilde{\ble}'(\w)$ at frequency $\w$ 
depends, to lowest order in $\ble$,  
only on $\tilde{\ble}(\w)$ at the same frequency $\w$ since the 
system is initially in equilibrium (hence invariant under time 
translations).
This implies that if the probe field has $N$ 
frequencies $\w_{1},\ldots,\w_{N}$ then the total missing energy $E-E'$ is 
the sum of the missing 
energies of $N$ independent PA experiments carried out with monochromatic beams of 
frequencies $\w_{1},\ldots,\w_{N}$.
The same is true for the energy absorbed by the sample:
$\tilde{\bld}_{p}(\w)$ 
depends only on $\tilde{\ble}(\w)$
since the probe-induced dipole moment is linear in $\ble$. 
Therefore for systems in equilibrium and to lowest order in the 
probing fields we can write
\be
\tilde{S}(\w)=
i \w \;\tilde{\ble}^{\ast}(\w)\cdot\tilde{\bld}_{p}(\w)
+c.c.
\label{swa}
\ee

The approaches to calculate the right hand side of Eq. 
(\ref{swa}) can be grouped into two classes.
In one (recently emerging) class one  
perturbs the system with an em field $\ble(t)$, calculates the 
time-dependent  dipole moment either by solving the 
Schr\"odinger/Liouville equation or by using other 
methods,\cite{biry.2000,sszl.2007,kb.2000,dl.2007,fva.2009,agm.2011,mtksl.2012,lpuls.2014,dbcwr.2013,nkyetal.2013,udroetal.2014}  
and then Fourier transforms it.
The other (more traditional) class avoids time-propagations and works directly in 
frequency space. To lowest order in $\ble$ the Kubo formula gives
\be
d_{p,i}(t)=\sum_{j}\int dt' \chi_{ij}
(t,t')e_{j}(t')
\label{kubod}
\ee
where $\chi$
is the (retarded) dipole-dipole response function.
For a system with Hamiltonian $\hat{H}$
in the ground state $|\Q_{g}\ket$ of energy $E_{g}$ we have
\bea
i\chi_{ij}(t,t')\!&=&\!\th(t-t')\bra\Q_{g}|e^{i\hat{H}t}
\hat{d}_{i}e^{-i\hat{H}(t-t')}\hat{d}_{j}e^{-i\hat{H}t'}-{\rm H.c.}|\Q_{g}\ket
\nn\\
\!&=&\!\th(t-t')\bra\Q_{g}|\hat{d}_{i}e^{-i(\hat{H}-E_{g})(t-t')}\hat{d}_{j}-{\rm H.c.}|\Q_{g}\ket
\label{eqrf}
\eea
where H.c. stands for ``hermitian conjugate'' and $\hat{d}_{i}$ is 
the $i$-th component of the dipole-moment operator. As expected the equilibrium 
response function $\chi$
depends on the time-difference 
only. Fourier transforming Eq. (\ref{kubod}) and inserting the result 
into Eq. (\ref{swa}) we get
\be
\tilde{S}(\w)=\w
\sum_{ij}
\tilde{e}^{\ast}_{i}(\w)\callL_{ij}(\w)\tilde{e}_{j}(\w),
\label{swachi}
\ee
with  $\callL_{ij}(\w)\equiv 
i\left[\tilde{\chi}_{ij}(\w)-\tilde{\chi}_{ji}^{\ast}(\w)\right]$. 
The response function $\tilde{\chi}(\w)$ can be calculated by several 
means 
without performing a time propagation.
From the Lehmann 
representation of $\chi$  it is easy to verify that $\callL$ is positive 
semidefinite for positive frequencies and negative semidefinite 
otherwise. Consequently $\tilde{S}$ is manifestly positive, in agreement 
with Eq. (\ref{dW}). 

\subsection{Energy approach: PA out of equilibrium}
\label{eneapp:nePAS}

In a typical TR-PA experiment  both the pump and probe 
 fields are very short 
(fs-as) laser pulses with a delay $\t$ between them. If 
$\t<0$  then the probe acts before the pump and we recover the PA 
spectra of equilibrium systems. On the other hand if $\t>0$ then 
$\tilde{S}(\w)$ acquires a 
dependence on $\t$. This dependence can be used to 
follow the evolution of the system in real time. 
However, for the physical interpretation of what we are actually 
following it is necessary to generalize the equilibrium PA theory to 
nonequilibrium situations.

Let the external electric field $\blE_{\rm ext}$ be the sum of the 
pump field $\blE$ and probe field $\ble$, i.e., 
$\blE_{\rm ext}=\blE+\ble$. In this case Eq. (\ref{DeltaE}) yields the 
{\em total} energy absorbed by the system. As the
experiment detects only the energy of the transmitted probe 
field the use of the energy approach for TR-PA is not straightforward.
One might argue that the energy absorbed  from the probe  
is given by Eq. (\ref{DeltaE}) in which 
$\blE_{\rm ext}\to\ble$
\be
E_{\rm abs}\stackrel{?}{=}
-\int dt \;\ble( t)\cdot \frac{d}{dt}\bld( t) .
\label{guess}
\ee
However, this formula cannot be always correct. 
Suppose that the pump field is also 
feeble and can be treated as a small perturbation. Then, the 
transmitted probe field $\tilde{\ble}'(\w)$ depends on 
$\tilde{\blE}(\w)+\tilde{\ble}(\w)$ (linear response theory). These 
fields are independent of position {\em 
inside} the sample. In a larger space, like that of 
the laboratory, they do depend on $\blr$ and this dependence
specifies the direction of propagation. Let $\tilde{\blE}(\blK\w)$ 
and $\tilde{\ble}(\blk\w)$ be the spatial Fourier transform of the pump 
and probe fields. For isotropic systems
$\tilde{\ble}'(\blk\w)$ depends only on $\tilde{\ble}(\blk\w)$
since $\tilde{\blE}(\blk\w)$ vanishes for $\blk$ parallel to the 
direction of propagation of the probe. This implies that 
the missing energy per unit frequency is independent of the pump, a 
conclusion which is 
not in agreement with Eq. (\ref{guess}). In fact, $E_{\rm abs}$ depends on 
 $\blE$ whenever the pump-induced
variation of $\bld$ is not orthogonal to $\ble$. 

To cure this problem we could write $\bld=\bld_{P}+\bld_{p}$, where 
$\bld_{P}$ is the value of the dipole moment when only the pump field 
is present
whereas 
$\bld_{p}$ is the probe-induced variation, and say that 
the missing probe energy is 
\be
E_{\rm abs}\stackrel{?}{=}
-\int dt \;\ble( t)\cdot \frac{d}{dt}\bld_{p}( t) .
\label{guess2}
\ee
This expression is by construction correct for perturbative pumps. For 
pumps of arbitrary strength Eq. (\ref{guess2}) cannot be proved or 
disproved using exclusively energy considerations. 
For the sake of the argument, however, let us assume that Eq. 
(\ref{guess2}) is the correct missing energy. 
There is still a conceptual problem to overcome if we are interested in  
the missing energy {\em per unit frequency}.
For strong pump fields the sample is in a nonstationary state and 
hence the transmitted probe field $\tilde{\ble}'(\w)$
 depends on the entire function $\tilde{\ble}$,
not only on the value
$\tilde{\ble}(\w)$ at the same frequency.
Thus the reasoning made below Eq.~(\ref{eabs2}) does not apply.
In particular if $\ble$  is monochromatic 
then $\ble'$ (as well as $\bld_{p}$) is, in general, not monochromatic. 
Consequently $\tilde{S} \propto |\tilde{\ble}|^{2}-|\tilde{\ble}'|^{2}$ is, in 
general, not monochromatic either. If we used the formula in Eq. 
(\ref{swa})  we 
would instead find that $\tilde{S}$ is peaked at only one frequency
since $\tilde{\ble}$ is peaked at only one frequency. 
To overcome these problems one has to abandon the 
energy approach
and calculate explicitly the transmitted probe field.

\subsection{Maxwell approach}

To overcome the difficulties of the energy approach we use the 
Maxwell equations to calculate explicitly the transmitted probe field. 
In a nonmagnetic medium the {\em total} electric field $\bcallE$, i.e., the sum 
of the external and induced field, satisfies the 
equation\cite{mukamel-book,loudon-book}
\be
\nabla^{2}\bcallE-\frac{1}{c^{2}}\frac{\de^{2}\bcallE}{\de t^{2}}=
-\frac{4\p}{c^{2}}\frac{\de\bra\blJ\ket}{\de t},
\label{poleq2}
\ee
where $\bra\blJ\ket(\blr t)$ is the macroscopic current density, i.e., the
spatial average of the current density over small volumes around 
$\blr$. In the derivation of Eq. (\ref{poleq2}) one uses that 
$\grad\times\grad\times \bcallE = 
\grad(\grad\cdot\bcallE)-\nabla^{2}\bcallE$ and that 
$\grad\cdot\bcallE=0$ since the 
sample is charge neutral (in a macroscopic sense).
Let $\hat{\blN}$ and $\hat{\bln}$ be 
the unit vectors along the propagation direction of the pump and 
probe fields respectively. In TR-PA experiments these vectors are not 
parallel for otherwise the detector would measure the transmitted 
pump intensity too. The time-dependence of the macroscopic current 
density arises when 
the pump and probe fields interact with the 
electrons in the sample. For transverse pump and probe fields and for 
isotropic systems 
$\bra\blJ\ket$ is the sum of transverse waves propagating 
along the directions $QK\hat{\blN}+qk\hat{\bln}$ 
with $Q$ and $q$ integers, and $K$ and $k$ the pump and probe wave 
numbers respectively. Consequently, the total electric field too is 
the sum of waves propagating along $QK\hat{\blN}+qk\hat{\bln}$, and Eq. 
(\ref{poleq2}) can be solved for each direction separately.

We define $\bcallE_{p}$ and 
$\bra\blJ\ket_{p}$ as the 
wave of the electric field and current density propagating toward the 
detector (hence $Q=0$ and $q=1$). 
The vectors $\bcallE_{p}$ and 
$\bra\blJ\ket_{p}$ depend on the spatial position $\blr$ 
only through $x=\hat{\bln}\cdot\blr$ (transverse fields) and are 
parallel to some unit vector $\bgve_{p}$ lying on the plane 
orthogonal to $\hat{\bln}$ (the generalization to multiple 
polarization is straightforward, see Section \ref{3levsec}): $\bcallE_{p}=\bgve_{p}\callE_{p}(x\,t)$ and 
$\bra\blJ\ket_{p}=\bgve_{p}J_{p}(x\,t)$.
Equation~(\ref{poleq2}) implies that 
\be
\frac{\de^{2}\callE_{p}}{\de 
x^{2}}-\frac{1}{c^{2}}\frac{\de^{2}\callE_{p}}{\de 
t^{2}}=-\frac{4\p}{c^{2}}\frac{\de J_{p}}{\de t},
\label{poleq4}
\ee
which is a one-dimensional wave equation that can be solved exactly without 
assuming slowly-varying probe envelops\cite{sypl.2011} or ultrathin 
samples.\cite{blm.2012} The electric field 
$\callE_{p}$ is 
the sum of an arbitrary solution $h(t-x/c)$ of the 
homogeneous equation and of an arbitrary special solution 
$s(x\,t)$: $\callE_{p}(x\,t)=h(t-x/c)+s(x\,t)$.
Without loss of generality we take the boundaries of the sample 
at $x=0$ and $x=L$. Let $e(t-x/c)$ be the amplitude of the incident probe field which at 
time $t=0$ is localized somewhere on the left of the sample.  
Imposing the boundary condition $\callE_{p}(x\,0)=e(-x/c)$ we then 
obtain
\be
\callE_{p}(x\,t)=e(t-x/c)-s(x-ct \,0)+s(x\,t).
\ee
The special solution $s(x\,t)$ is found by inverting the one-dimensional 
d'Alambertian 
$\oblong\equiv \frac{\de^{2}}{\de 
x^{2}}-\frac{1}{c^{2}}\frac{\de^{2}}{\de t^{2}}$. 
The Green's function 
$G(x\,t)$ solution of $\oblong G(x\,t)=\d(x)\d(t)$ is
\be
G(x\,t)=-\frac{c}{2}\,\th(t)\chi_{[-ct,ct]}(x),
\ee
where $\chi_{[a,b]}(x)=1$ if $x\in (a,b)$ and zero otherwise. 
Therefore the special solution reads
\be
s(x\,t)=\frac{2\pi}{c}\int_{-\iif}^{t}dt'\int_{-x-c(t-t')}^{x+c(t-t')}dx'\,
\frac{\de J_{p}(x' t')}{\de t'}.
\label{partsol}
\ee
Without any loss of generality we can choose the time $t=0$ as the time 
before which nor the pump and neither the probe have reached the 
sample. Then $J_{p}=0$ for $t<0$ and hence $s(x\,0)=0$ for 
all $x$. We conclude that
\be
\callE_{p}(x\,t)=e(t-x/c)+s(x\,t),
\label{e(xt)}
\ee
with $s$ given in Eq.~(\ref{partsol}).

We are interested in the electric field on the right
of the sample, i.e., in $x=L$, since this is the detected field. 
Let us therefore evaluate Eq. (\ref{partsol}) in $x=L$. 
Taking into account that $J_{p}(x't')$ is 
nonvanishing 
only for $x'\in(0,L)$ and $t'>0$ we have
\bea
s(L\,t)
&=&\frac{2\p}{c}\int_{0}^{t}dt'\int_{0}^{L}dx'\,
\frac{\de J_{p}(x't')}{\de t'}
\nn\\
&=&\frac{2\p}{\callS c}\bgve_{p}\cdot \int_{V} d\blr \,
\bra \blJ\ket_{p}(\blr t),
\label{partsol2}
\eea
where in the second line we integrated over the volume 
$V=\callS L$ of the sample. Using again the identity $\bra \blJ\ket_{p}=\grad(\blr\cdot\bra \blJ\ket_{p})
-\blr(\grad\cdot\bra \blJ\ket_{p})$ and extending the integral over all 
space (outside $V$ the current density vanishes) we can rewrite Eq. 
(\ref{partsol2}) as $s(L\,t)=-\frac{2\p}{\callS c}\bgve_{p}\cdot\int d\blr 
\,\blr\,(\grad\cdot\bra \blJ\ket_{p})$. 
Substituting this result into Eq~(\ref{e(xt)}) and taking into 
account the continuity equation 
$\grad\cdot\bra \blJ\ket_{p}=-\de\bra n\ket_{p}/\de t$, where $\bra 
n\ket_{p}$ is the macroscopic probe-induced change of the electronic 
density propagating along $\hat{\bln}$, 
we eventually obtain the transmitted electric field 
\be
\callE_{p}(L\,t)=
e(t)+\frac{2\p}{\callS c}\bgve_{p}\cdot \frac{d}{dt}\int d\blr \,\blr 
\,\bra n\ket_{p}(\blr t),
\label{epinL}
\ee
where we discarded the delay $L/c$ in the first term on the right 
hand side. The volume integral is the probe-induced dipole moment 
propagating along $\hat{\bln}$. In general this is not the same as 
the full probe-induced dipole moment $\bld_{p}$ defined above 
Eq.~(\ref{guess2}) since, to lowest order in the probe field,
$\bld_{p}$ is the sum of waves propagating along 
$k\hat{\bln}+QK\hat{\blN}$. Although it is reasonable to expect that 
the wave propagating along $\hat{\bln}$ (i.e., with $Q=0$) has the largest amplitude,
it is important to bear in mind this conceptual difference. In fact,
in equilibrium PA experiments $\bld_{p}$ and $\int d\blr \,\blr 
\,\bra n\ket_{p}$ are the same due to the absence of the pump.
For not introducing too many symbols we {\em redefine} 
$\bld_{p}\equiv \int d\blr \,\blr 
\,\bra n\ket_{p}$. Then, by definition, $\bld_{p}$ is parallel to 
$\bgve_{p}$ and we can cast Eq.~(\ref{epinL}) in vector 
notation as
\be
\ble'(t)=\ble(t)+\frac{2\p}{\callS c}
\frac{d}{dt}\bld_{p}(t).
\label{main}
\ee
Equation (\ref{main}) relates the transmitted probe field to the 
quantum-mechanical average of the probe-induced dipole moment, and it  
represents the fundamental bridge between theory and experiment.
The result has been derived without assuming that the 
wavelength of the incident field is much larger than the 
longitudinal dimension of the sample (for thick samples $\callE_{p}$ 
can be substantially different from $e$ and the quantum electron dynamics 
should be coupled to the Maxwell equations). 
Equation~(\ref{main}) can be used to calculate the missing energy per unit 
frequency of pump-driven systems. Noteworthy Eq. 
(\ref{main}) is valid for positive and negative 
delays between pump and probe, as well as for situations in 
which pump and probe overlap in time or even for more exotic 
situations in which the pump is entirely contained in the time-window 
of the probe. 

\section{Nonequilibrium response function}
\label{nechisec}

From the definition in Eq.~(\ref{dW}) the spectrum of a (equilibrium 
or nonequilibrium) PA experiment is given by
\be
\tilde{S}(\w)=\callS\frac{c}{2\p}\left(
|\tilde{\ble}(\w)|^{2}-|\tilde{\ble}'(\w)|^{2}\right).
\label{dw2}
\ee
Since $\ble=\bgve_{p}e$ and $\bld_{p}=\bgve_{p}d_{p}$ are both 
parallel to $\bgve_{p}$, so it is $\ble'=\bgve_{p}e'$.
Then, the Fourier transform of Eq. (\ref{main}) yields
$\tilde{e}'(\w)=\tilde{e}(\w)-i\frac{2\p}{\callS c}
\w\,\tilde{d}_{p}(\w)$, and the spectrum in Eq.~(\ref{dw2}) can be rewritten as
\bea
\tilde{S}(\w)=-2\Im\left(\w\,\tilde{e}^{\ast}(\w)\tilde{d}_{p}(\w)\right)
-\frac{2\p}{\callS c}\left|\w\,\tilde{d}_{p}(\w)\right|^{2}.
\label{nedw}
\eea

At the end of Section \ref{eneapp:nePAS} we 
criticized the energy approach since it predicts a single-peak 
spectrum for monochromatic probes. Let us analyze 
Eq. (\ref{nedw}) for the same case.
For monochromatic probes of frequency $\w_{0}$ the 
first term in Eq. (\ref{nedw}) vanishes for $\w\neq\w_{0}$ whereas the 
second term is nonvanishing at the same frequencies of the transmitted 
probe field, see Eq.~(\ref{main}), in 
agreement with the discussion at the end of Section \ref{eneapp:nePAS}. 
The quadratic term in the dipole moment is usually 
discarded in equilibrium PA calculations since $d_{p}$ and $e$ oscillate
at the same frequencies, and typically
$|\tilde{e}(\w)|\gg (2\p/\callS c)|\w\tilde{d}_{p}(\w)|$. 
If we discard the last term in Eq. (\ref{nedw}) then we recover the spectrum 
of Eq. (\ref{swa}) of the energy approach.

For the physical interpretation of nonequilibrium PA spectra it is 
crucial to understand the physics contained in the nonequilibrium 
dipole-dipole response function. In fact, 
$d_{p}$ can be 
calculated from the scalar version of the Kubo formula in Eq. 
(\ref{kubod}), i.e.,\cite{note2}  
\be
d_{p}(t)=\int dt'\chi(t,t')\callE_{p}(t').
\label{scalarkubo}
\ee
Here the 
scalar dipole-dipole response function $\chi$ is defined according to 
$\chi=\sum_{ij}\ve_{p,i}\chi_{ij}\ve_{p,j}$ and $\callE_{p}$ is the 
{\em total} electric field ($\callE_{p}\simeq e$ if the induced field 
is small).
Unfortunately, for pump-driven systems  
a Lehmann-like formula for $\chi$ does 
not exist  due to the presence of a strong time-dependent 
perturbation  in the Hamiltonian. By introducing
the evolution operator $\hat{\callU}(t,t')$  from $t'$ to 
$t$ of the system without the probe, 
the nonequilibrium dipole-dipole response function reads
\bea
i\chi(t,t')\!&=&\!\th(t-t')
\nn\\
\!&\times&\!
\bra\Q_{g}|\hat{\callU}(t_{0},t)
\hat{d}\,\hat{\callU}(t,t')\hat{d}\,\hat{\callU}(t',t_{0})-{\rm H.c.}|\Q_{g}\ket
\nn\\
\label{nerf1}
\eea
where $\hat{d}=\bgve_{p}\cdot\hat{\bld}$ and
$t_{0}$ is any time earlier than the switch-on time of the pump and probe fields.
As anticipated the nonequilibrium $\chi$  depends on $t$ and $t'$ 
separately.
It is clear from Eq. (\ref{nerf1}) that $\chi$ does not have a simple 
representation in terms of the many-body eigenstates and 
eigenenergies of the unperturbed system.
It is also easy to verify that Eq. (\ref{nerf1}) agrees with
Eq.~(\ref{eqrf}) in the absence of the pump.

As a final remark before presenting some exact properties of $\chi$, 
we observe that
in equilibrium, see Eq. (\ref{swachi}), the ratio $\tilde{S}(\w)
/|\tilde{e}(\w)|^{2}$ is independent of the probe field, i.e.,  it is 
an intrinsic property of the sample. 
This is not true in nonequilibrium, even if we discard the last 
term in Eq.~(\ref{nedw}) and approximate $\callE\simeq e$.
The physical interpretation of nonequilibrium PA spectra  
cannot leave out of consideration the shape and duration of the probe, 
and the relative delay between pump and probe. In the next 
sections we discuss two relevant situations for interpreting
the outcome of a TR-PA experiment.

\section{Nonoverlapping pump and probe}
\label{nonoverlapsec}

Let us consider the case of a probe pulse acting {\em after} the pump 
pulse. We take the time origin $t=0$ as the switch-on time of the 
probe. Then the pump acts at some time $t=-\t<0$. 
For $t>0$ 
the probe-induced variation of the dipole moment can be calculated
from Eq. (\ref{scalarkubo}) with lower integration limit $t'=0$. As we 
only need $\chi$ for $t,t'>0$ we have
$\hat{\callU}(t_{0},t)=\hat{\callU}(t_{0},0)e^{i\hat{H}t}$ and 
similarly 
$\hat{\callU}(t',t_{0})=e^{-i\hat{H}t'}\hat{\callU}(0,t_{0})$, with 
$\hat{H}$ the unpertubed Hamiltonian of the system. Defining 
$|\Q\ket\equiv\hat{\callU}(0,t_{0})|\Q_{g}\ket$ as the quantum state of the 
system at time $t=0$, the response function in Eq. (\ref{nerf1}) becomes
\be
i\chi(t,t')=\th(t-t')\bra\Q|e^{i\hat{H}t}
\hat{d}\,e^{-i\hat{H}(t-t')}\hat{d}\,e^{-i\hat{H}t'}-{\rm H.c.}|\Q\ket
\label{nerf2}
\ee
which closely resembles the equilibrium response function of Eq. 
(\ref{eqrf}). In fact, without pump fields 
$|\Q\ket=e^{iE_{g}t_{0}}|\Q_{g}\ket$ and Eq. (\ref{nerf2}) reduces to
the equilibrium response function. We emphasize that in the presence 
of pump fields Eq. 
(\ref{nerf2}) is valid only for $t,t'>0$.

We expand the quantum state 
$|\Q\ket=\sum_{\a}c_{\a}|\Q_{\a}\ket$ in terms 
of the many-body eigenstates $|\Q_{\a}\ket$ of $\hat{H}$ with 
eigenenergy $E_{\a}$. The coefficients 
$c_{\a}=\bar{c}_{\a}e^{-iE_{\a}\t}$ depend on 
the delay $\t$ between the pump  and the 
probe, $\bar{c}_{\a}$ being the expansion coefficients of the state 
of the system at the end of 
the pump.
Inserting the 
expansion in Eq. (\ref{nerf2}) and using the completeness relation 
$\sum_{\g}|\Q_{\g}\ket\bra\Q_{\g}|=1$ we find
\be
i\chi(t,t')=\th(t-t')\sum_{\a\b\g}c^{\ast}_{\a}c_{\b}e^{i\W_{\a\g}t}
e^{i\W_{\g\b}t'}d_{\a\g}d_{\g\b}-c.c.
\label{nerf3}
\ee
where we defined the dipole matrix elements 
$d_{\a\g}=\bra\Q_{\a}|\hat{d}|\Q_{\g}\ket$ and the energy 
differences $\W_{\a\g}=E_{\a}-E_{\g}$. 
In the following we use this result to study the outcome of a PA experiment in 
two limiting cases, i.e., an ultrashort probe and a monochromatic 
probe.

\subsection{Ultra-short probe}
\label{ultrashortprobesec}

The probe fields used in TR-PA experiments are 
ultrashort laser pulses. For optically thin samples
the em field generated by the 
probe-induced dipole moment is negligible and
does not substantially affect the quantum evolution 
of the system. Therefore, we can calculate $d_{p}$ from Eq. 
(\ref{scalarkubo}) with $\callE=e$. For a delta-like probe 
$e(t)=e_{0}\d(t)$, hence 
$\tilde{e}(\w)=e_{0}$, we find
\bea
d_{p}(t)=-ie_{0}
\sum_{\a\b\g}c^{\ast}_{\a}c_{\b}e^{i\W_{\a\g}t}
d_{\a\g}d_{\g\b}+c.c.
\label{ned5}
\eea
where we used the explicit form of the response function in Eq. 
(\ref{nerf3}). Fourier transforming this result we obtain the  spectrum
\bea
\tilde{S}(\w)&=&-2\w e_{0}^{2}
\sum_{\a\b\g}\Im\big[e^{i\W_{\a\b}\t}\,\bar{c}^{\ast}_{\a}\bar{c}_{\b}d_{\a\g}d_{\g\b}
\nn\\
&\times&
\big(\frac{1}{\w-\W_{\g\a}+i\eta}-\frac{1}{\w+\W_{\g\b}+i\eta}\big)\big]
\label{deltaspectrum}
\eea
where $\eta$ is a positive infinitesimal and we discarded the 
quadratic term in $d_{p}$ (thin samples). In Eq.~(\ref{deltaspectrum})
the dependence on the delay $\t$ enters exclusively 
through the phase factors and, consequently, it is only responsible for 
modulating the amplitude of the absorption peaks. The position of the peaks is 
instead an intrinsic property of the unperturbed system. Thus,  a 
change in the peak-position (discrete spectrum) or in the onset of a 
continuum (continuum 
spectrum) due to $\t$ should not be attributed to a change of the 
many-body energies but to a redistribution of the spectral weights.

Let us discuss Eq.~(\ref{deltaspectrum}) in some detail.
For a system in equilibrium in the ground state (no pump) $c_{\a}=1$ for $\a=g$ and 
$c_{\a}=0$ otherwise, and the spectrum reduces to
\be
\tilde{S}(\w)=2\p\w e_{0}^{2}
\sum_{\g}|d_{g\g}|^{2}
\left(\d(\w-\W_{\g g})-\d(\w+\W_{\g g})\right).
\label{deltaspectrumg}
\ee
Since $E_{\g}>E_{g}$ the spectrum 
is nonnegative, in agreement with Eq.~(\ref{dW}).
In particular the height of the peak at some frequency $\w_{0}$ is given by 
$h_{g}=2\p|\w_{0}|e_{0}^{2}\sum_{\g:\W_{\g g}=|\w_{0}|}|d_{g\g}|^{2}\geq 0$.
It is also interesting to consider the hypothetical situation of a pump pulse which 
brings the system from the ground state to an excited 
state $|\Q\ket=|\Q_{x}\ket$ with energy $E_{x}$. 
As the system is stationary this is the simplest example  of a nonequilibrium 
situation. In the stationary case we have $c_{\a}=1$ for $\a=x$ 
and $c_{\a}=0$ 
otherwise, and hence the spectrum is again given by 
Eq.~(\ref{deltaspectrumg}) with the only difference that the 
subscript ``$g$'' is replaced by the subscript ``$x$''. Since $E_{x}$ 
is not the lowest energy the positivity of the spectrum is no longer  
guaranteed. In fact,
the height of the peak at frequency $\w_{0}$ is 
\be
h_{x}=2\p\w_{0}e_{0}^{2}\,\big(\sum_{\g:\W_{\g x}= \w_{0}}|d_{x\g}|^{2}
-\sum_{\g:\W_{\g x}= -\w_{0}} |d_{x\g}|^{2}\big)
\label{hstatstate}
\ee
which can be either positive or 
negative. 
 The sign is positive if the absorption rate is larger than 
the rate for stimulated emission and negative otherwise.
We observe that in the stationary case the spectrum is 
{\em independent} of the delay.

The most general situation is a system in 
a nonstationary state. From Eq.~(\ref{deltaspectrum})
the peak intensity at some frequency $\w_{0}$ reads
\be
h=2\p\w_{0}e_{0}^{2}\sum_{\a\b}
\sum_{\pm}\pm\!\!\sum_{\g:\W_{\g\a}= \pm\w_{0}}
\Re\left[e^{i\W_{\a\b}\t}\,\bar{c}^{\ast}_{\a}\bar{c}_{\b}d_{\a\g}d_{\g\b}\right],
\label{ufh}
\ee
where we introduced the short-hand notation 
$\sum_{\pm}A_{\pm}=A_{+}+A_{-}$, with $A$ an arbitrary mathematical 
expression.
If $|\Q\ket$ is a superposition of degenerate 
eigenstates, hence $\bar{c}_{\a}\neq 0$ for $E_{\a}=E$ and  $\bar{c}_{\a}= 
0$ otherwise, then the system is stationary and the height is independent 
of the delay. The dependence on $\t$ is manifest only for $|\Q\ket$ 
a superposition of 
{\em nondegenerate} eigenstates. The simplest example is 
a system in a superposition of two eigenstates with energy
$E_{a}$ and $E_{b}$, real coefficients $\bar{c}_{a}$ and 
$\bar{c}_{b}$ and real dipole matrix 
elements. In this case Eq.~(\ref{ufh}) yields 
$h=h_{a}+h_{b}+h_{ab}\cos(\W_{ab}\t)$ where $h_{x=a,b}$ is defined as in Eq. 
(\ref{hstatstate}) and  
\be
h_{ab}= 2\p\w_{0}e_{0}^{2}\bar{c}_{a}\bar{c}_{b}
\sum_{\a=a,b}\sum_{\pm}\pm\!\!\sum_{\g:\W_{\g 
\a}=\pm\w_{0}}d_{a\g}d_{b\g}.
\nn
\ee
The spectral fingerprint of a nonstationary system 
is the modulation of the peak intensities with $\t$. These coherent 
oscillations have been first observed in 
Ref.~\onlinecite{glwsetal.2010}.

\subsection{Monochromatic probe}
\label{monoprobesec}

The induced electric field of thick samples is not negligible and 
can last much longer than the external probe 
pulse. In this case the quantum evolution of the system should be 
coupled to the Maxwell equations to determine the total field
self-consistently.\cite{mukamel-book,zag.1995,bsy.2007,gbts.2011,cl2.2012,csg.2012,cl.2013,wcsg.2013,pbbmnl.2013,oyi.2009,yssob.2012,wlbsetal.2014,ds.2014} 
For a typical sub-as pulse $e(t)$ centered around a resonant frequency $\w_{0}$ the 
total electric field $\callE(t)$ is dominated by oscillations of 
frequency $\w_{0}$ decaying over the same time-scale of the induced 
dipole moment (in atomic gases  this time-scale can be as long as  hundreds of fs).
Let us explore the outcome of a TR-PA experiment for  
a total field of the form, e.g., 
$\callE(t)=\callE_{0}\th(t)\sin (\w_{0}t)$, $\w_{0}>0$. Taking into account Eq. 
(\ref{nerf3}) we find
\be
d_{p}(t)=
\frac{i}{2}\callE_{0}\sum_{\a\b\g}\sum_{\pm}c^{\ast}_{\a}c_{b}d_{\a\g}d_{\g\b}
\frac{e^{i(\W_{\a\b}\pm\w_{0})t}-e^{i\W_{\a\g}t}}{\pm (\W_{\g\b}\pm\w_{0})}
+c.c.
\label{ned2}
\ee
If $|\Q\ket=|\Q_{x}\ket$ (stationary system) then $c_{\a}=\d_{\a 
x}$ and
the dominant contributions in Eq.~(\ref{ned2}) come from  
eigenstates with energy 
$E_{\g}=E_{x}+\w_{0}$ in the ``$-$'' sum  and from eigenstates with energy 
$E_{\g}=E_{x}-\w_{0}$ in the ``$+$'' sum. Therefore Eq.~(\ref{ned2}) 
is well approximated by
\be
d_{p}(t)=\callE_{0}\,t\cos(\w_{0}t)\sum_{\pm}\pm\sum_{\g:\W_{\g 
x}=\pm\w_{0}}|d_{x\g}|^{2}.
\label{edss}
\ee
As expected the dipole moment oscillates at the same frequency of the 
electric field. Unlike the spectrum of an ultrashort probe, in the 
monochromatic case $\tilde{S}(\w)$
has at most one peak. For $x=g$ (ground state) the ``$-$'' sum vanishes since 
$\W_{\g g}>0$, and we recover 
the well know physical interpretation of 
equilibrium PA experiments: peaks in $\tilde{S}(\w)$
occur in correspondence of the energy 
of a charge neutral excitation. This remains true for $x\neq g$
but the sign of the oscillation amplitude can be either positive 
or negative.
Notice that the oscillation amplitude is proportional 
to $h_{x}$ in Eq.~(\ref{hstatstate}) and it is independent of the delay.

In the nonstationary case $|\Q\ket$ is a superposition of 
nondegenerate eigenstates.
For a fixed $\b$ in Eq. 
(\ref{ned2}) the dominant contributions
come from eigenstates with energy 
$E_{\g}=E_{\b}+\w_{0}$ in the ``$-$'' sum  and from eigenstates with energy 
$E_{\g}=E_{\b}-\w_{0}$ in the ``$+$'' sum. Writing 
$\W_{\a\g}=\W_{\a\b}+\W_{\b\g}$ it is straightforward to show that
\bea
&d_{p}(t)&\!\!=\callE_{0}\frac{t}{2}
\sum_{\a\b}\sum_{\pm}\pm \,c^{\ast}_{\a}c_{\b}e^{i\W_{\a\b}t}
\nn\\
&\times&\!\!\!\!\!\!\!
\big(e^{\mp 
i\w_{0}t}\!\!\sum_{\g:\W_{\g\b}=\pm\w_{0}}d_{\a\g}d_{\g\b}
+e^{\pm i\w_{0}t}\!\!\sum_{\g:\W_{\g\a}=\pm\w_{0}}
 d_{\a\g}d_{\g\b}\big).\quad\quad
\label{edns}
\eea
As anticipated below Eq.~(\ref{nedw}) $d_{p}(t)$ is not monochromatic in a 
nonstationary situation, the frequencies of the oscillations being 
$\w_{0}\pm\W_{\a\b}$. Can these extra frequencies be seen in the 
TR-PA spectrum? The answer is affirmative since $\tilde{e}(\w)$ is a broad function 
centered in $\w_{0}$ and hence for $|\W_{\a\b}|$ not too large 
$\Im[\tilde{e}^{\ast}(\w_{0}\pm\W_{\a\b})\tilde{d}_{p}(\w_{0}\pm\W_{\a\b})]$ is 
nonvanishing. Furthermore, the 
induced electric field is sizable and hence the second term in the 
right hand side of Eq. 
(\ref{nedw}) cannot be discarded. 

The dipole moment in Eq.~(\ref{edns}) is substantially different from 
the ultrafast probe-induced $d_{p}$ of Eq.~(\ref{ned5}). In order to
appreciate the difference we calculate the amplitude of the 
dipole oscillation of frequency $\w_{0}$ and compare it with the height $h$ 
in Eq.~(\ref{ufh}). By restricting the sum over $\a,\b$ to states 
with $\W_{\a\b}=0$ in Eq. (\ref{edns}), we obtain the harmonics 
$d_{p,\w_{0}}(t)$ with frequency $\pm\w_{0}$
\be
d_{p,\w_{0}}(t)=\callE_{0}\,t\cos(\w_{0}t)
\sum_{\a\b:\W_{\a\b}=0}\bar{c}^{\ast}_{\a}\bar{c}_{\b}
\sum_{\pm}\pm\!\!\sum_{\g:\W_{\g\a}=\pm\w_{0}}d_{\a\g}d_{\g\b}.
\label{monoharm}
\ee
The main difference between the amplitude of $t\cos(\w_{0}t)$ in 
Eq.~(\ref{monoharm}) and the 
peak height in Eq. (\ref{ufh}) is that in the former we have a 
constrained sum over $\a,\b$. Consequently no 
coherent oscillations as a function of
the delay $\t$ are observed in the TR-PA spectrum around frequency 
$\w_{0}$, in agreement with recent experimental findings in 
Ref.~\onlinecite{pbbmnl.2013}.

\section{Overlapping pump and probe}
\label{overlapsec}

In the overlapping regime the difficulty in extracting 
physical information from the 
nonequilibrium response function is due to the presence of the pump 
in the evolution operator. 
Nevertheless, some analytic progress can still be 
made for the relevant case of ultrashort probes. If the pump is active for a long 
enough time before and after the probe then we can approximate it 
with an everlasting field. 
In this Section we study the 
nonequilibrium response function of systems driven out of 
equilibrium by a strong periodic pump field. As we shall see in Section 
\ref{numsec} this analysis will help the interpretation of 
TR-PA spectra. 

Let us consider a periodic Hamiltonian 
$\hat{H}(t)=\hat{H}(t+T_{P})=\sum_{n}e^{in\w_{P} t}\hat{H}^{(n)}$, with 
$\w_{P}=2\p/T_{P}$. According to the Floquet theorem the evolution operator 
can be expanded as\cite{s.1965}
\be
\callU(t,t')=\sum_{\a}e^{-iQ_{\a}(t-t')}|\Q_{\a}(t)\ket\bra\Q_{\a}(t')|.
\label{evolflocquet}
\ee
In this equation 
$|\Q_{\a}(t)\ket=|\Q_{\a}(t+T_{P})\ket=\sum_{n}e^{in\w_{P} 
t}|\Q_{\a}^{(n)}\ket$ are the quasi-eigenstates and $Q_{\a}$ are the 
quasi-energies. They are found by solving the  
Floquet eigenvalue problem 
$\sum_{k}\hat{\callH}_{F,nk}|\Q_{\a}^{(k)}\ket=Q_{\a}|\Q_{\a}^{(n)}\ket$
with $\hat{\callH}_{F,nk}\equiv \hat{H}^{(n-k)}+n\w_{P}\d_{nk}$.
It is easy to show that if $\big\{Q_{\a},|\Q_{\a}^{(n)}\ket\big\}$ is a 
solution then $\big\{Q'_{\a}=Q_{\a}+m\w_{P},|\Q_{\a}^{\prime (n)}\ket=|\Q_{\a}^{(n-m)}\ket\big\}$ is a 
solution too. These two solutions, however, are not independent since 
$e^{-iQ_{\a}t}|\Q_{\a}(t)\ket=e^{-iQ'_{\a}t}|\Q'_{\a}(t)\ket$.
In Eq.~(\ref{evolflocquet}) the sum over $\a$ is restricted to 
independent solutions. For 
time-independent Hamiltonians ($\hat{H}^{(n)}=0$ for all $n\neq 0$) the independent solutions reduce to the 
eigenvalues $E_{\a}$ and eigenvectors $|\Q_{\a}\ket$ of 
$\hat{H}=\hat{H}^{(0)}$.

We expand the ground state 
$|\Q_{g}\ket=\sum_{\a}b_{\a}e^{-iQ_{\a}t_{0}}|\Q_{\a}(t_{0})\ket$ 
in quasi-eigenstates and
insert Eq.~(\ref{evolflocquet}) into Eq.~(\ref{nerf1}) to derive the following Lehmann-like 
representation of the nonequilibrium response function
\be
i\chi(t,t')=\th(t-t')\sum_{\a\b\g}b^{\ast}_{\a}b_{\b}e^{i\W_{\a\g}t}
e^{i\W_{\g\b}t'}d_{\a\g}(t)d_{\g\b}(t')-c.c.
\label{nerf3f}
\ee
In this formula $\W_{\a\b}=Q_{\a}-Q_{\b}$ are the quasi-energy differences and 
$d_{\a\b}(t)=\bra\Q_{\a}(t)|\hat{d}|\Q_{\b}(t)\ket$ are the 
time-periodic dipole 
matrix elements in the quasi-eigenstate basis. 

Let us compare the response function {\em 
during} the action of the pump, Eq. (\ref{nerf3f}),  with the 
response function {\em after} the action of the pump, Eq. (\ref{nerf3}). 
Unlike the coefficients $c_{\a}$ of the expansion of $|\Q\ket$ 
(this is the state of the system after a time $\t$ from the 
switch-off time of the pump) the 
coefficients $b_{\a}$ of the expansion of the ground 
state $|\Q_{g}\ket$ are independent of the delay. 
Bearing this difference in mind we can repeat step by step the derivations of 
Section \ref{ultrashortprobesec} and \ref{monoprobesec} with 
$E_{\a}\to Q_{\a}$ and $d_{\a\b}\to 
d_{\a\b}(t)=\sum_{n}e^{in\w_{P}t}d_{\a\b}^{(n)}$. 
We then conclude that the absorption regions occur in correspondence 
of the quasi-energy differences and of their replicas (shifted by 
integer multiples of $\w_{P}$). 

It would be valuable to relate the quasi-energies to 
the period and intensity of the pump field. In general, however, this relation 
is extremely complicated. In the following we discuss the special 
case of monochromatic pumps, which is also relevant when 
treating other periodic fields in the rotating wave 
approximation.\cite{Allen-book,Meystre-book,Boyd-book,Cohen-Tannoudji-book}

\subsection{Monochromatic Pumps}

For a monochromatic pump the time-dependent light-matter 
interaction Hamiltonian has the general form $\hat{H}_{P}(t)=\hat{P}e^{-i\w_{P} 
t}+\hat{P}^{\dagger}e^{i\w_{P} t}$. Hence $\hat{H}^{(0)}=\hat{H}$ 
is the Hamiltonian of the unperturbed system, 
$\hat{H}^{(-1)}=\hat{P}$, $\hat{H}^{(1)}=\hat{P}^{\dag}$ and 
$\hat{H}^{(n)}=0$ for all $|n|>1$. Then the Floquet operator 
reads
\be
\hat{\callH}_{F}=\left( 
\begin{array}{cccccccc}
 & \vdots & \vdots & \vdots & \vdots & \vdots &  \\
\cdots & \hat{P}^{\dagger} & \hat{H} - \w_{P}  &  
\hat{P} & 0 & 0 & \cdots \\
\cdots &0 &\hat{P}^{\dagger} & \hat{H} & 
\hat{P} & 0 & \cdots \\
\cdots & 0 &  0 &  \hat{P}^{\dagger} & \hat{H} + \w_{P}   &  \hat{P} & \cdots \\
 & \vdots & \vdots & \vdots & \vdots & \vdots &  
\end{array} \right)  \, .
\ee
The operator $\hat{\callH}_{F}$ acts on the direct 
sum of infinite Fock spaces.
The tridiagonal block structure allows for reducing the 
dimensionality of the Floquet eigenvalue problem.  With a standard 
embedding technique it is easy to show that the quasi-energies $Q_{\a}$ and the 
zero-th harmonic $|\Q_{\a}^{(0)}\ket$ of the quasi-eigenstates 
are solutions of $\hat{H}_{\rm eff}(Q)|\Q^{(0)}\ket=Q|\Q^{(0)}\ket$ 
where
\begin{widetext}
\be
\hat{H}_{\rm eff}(Q) = \hat{H} + 
\hat{P}^{\dagger}\frac{1}{Q- \hat{H} -\w_{P} -\hat{P}^{\dagger} 
\frac{1}{Q- \hat{H} -2\w_{P} -\hat{P}^{\dagger} \frac{1}{Q- 
\hat{H} -3\w_{P} -\cdots} \hat{P} } \hat{P}   } 
\hat{P} +
\hat{P}\frac{1}{Q- \hat{H} +\w_{P} -\hat{P}
\frac{1}{Q- \hat{H} +2\w_{P} -\hat{P} \frac{1}{Q- 
\hat{H} +3\w_{P} -\cdots} \hat{P}^{\dagger} } \hat{P}^{\dagger}   } 
\hat{P}^{\dagger}  \, .
\ee
\end{widetext}
In the large-$\w_{P}$ limit the leading contribution
is $\hat{H}_{\rm eff}(Q) = \hat{H}+\frac{1}{\w_{P}}[\hat{P},\hat{P}^{\dag}]$, 
which can be diagonalized to address the high-energy spectral features.\cite{kobfd.2011,lkzs.2014} 

The Floquet eigenvalue problem simplifies considerably if 
we retain only matrix elements $P_{\a\b}\equiv\bra\Q_{\a}|\hat{P}|\Q_{\b}\ket$ with 
$E_{\a}-E_{\b}\simeq \w_{P}$, and if the subsets of indices $\a$ and $\b$ are 
disjoint. In this case the Fock space can be divided into two subspaces $A$ 
and $B$ with the property that 
$\hat{P}=\sum_{\a\in A,\b\in B}P_{\a\b}|\Q_{\a}\ket\bra\Q_{\b}|$. 
We write a state $|\Q\ket$ in Fock space  as 
$|\Q_{A}\ket+|\Q_{B}\ket$ with $|\Q_{X}\ket=\sum_{\c\in 
X}|\Q_{\c}\ket\bra\Q_{\c}|\Q\ket$, $X=A,B$. Similarly, 
we split the total Hamiltonian $\hat{H}=\hat{H}_{A}+\hat{H}_{B}$ into 
the sum of an operator 
$\hat{H}_{A}=\sum_{\a\in A}E_{\a}|\Q_{\a}\ket\bra\Q_{\a}|$ acting on 
subspace $A$ and an operator 
$\hat{H}_{B}=\sum_{\b\in B}E_{\b}|\Q_{\b}\ket\bra\Q_{\b}|$ acting on 
subspace $B$.
It is straightforward to verify that the Floquet eigenvalue problem 
decouples into pairs of equivalent equations involving two consecutive  
blocks. Choosing for instance the blocks with $n=0$ and $n=1$ we find
\be
\left( 
 \begin{array}{cc}
 \hat{H}_{A}   & \hat{P}  \\
  \hat{P}^{\dagger}   & \hat{H}_{B} + \w_{P}
  \end{array} \right) 
  	 \left(
\begin{array}{c}
|\Q_{\c A}^{(0)}\ket \\
|\Q_{\c B}^{(1)}\ket 
\end{array}
\right)
=Q_{\c}
 \left(
\begin{array}{c}
|\Q_{\c A}^{(0)}\ket \\
|\Q_{\c B}^{(1)}\ket 
\end{array}
\right) \, .
\label{smallmatrix}
\ee
The $2\times 2$ matrix on the left hand 
side is known as the 
Rabi operator.
From the solutions of Eq. (\ref{smallmatrix}) we can construct the 
full set of quasi-eigenstates according to 
$|\Q_{\c}(t)\ket=|\Q_{\c 
A}^{(0)}\ket+e^{i\w_{P}t}|\Q_{\c B}^{(1)}\ket$.\cite{note1} Thus the 
quasi-eigenstates  contain only a single replica. 
Notice that in the absence of pump fields the solutions are either
$Q_{\c}=E_{\c}$, $|\Q_{\c A}^{(0)}\ket=|\Q_{\c}\ket$ and $|\Q_{\c 
B}^{(1)}\ket=0$ or
$Q_{\c}=E_{\c}+\w_{P}$, $|\Q_{\c A}^{(0)}\ket=0$ and $|\Q_{\c 
B}^{(1)}\ket=|\Q_{\c}\ket$.

The single replica of the quasi-eigenstates reflects into 
a single replica of the time-dependent dipole matrix elements. In fact, the given form of the operator 
$\hat{P}$ implies that the dipole operator $\hat{d}$ couples states 
in subspace $A$ to states in subspace $B$ and viceversa. Therefore
\be
d_{\a\b}(t)=e^{i\w_{P}t}\bra\Q^{(0)}_{\a A}|
\hat{d}|\Q^{(1)}_{\b B}\ket
+e^{-i\w_{P}t}\bra\Q^{(1)}_{\a B}|
\hat{d}|\Q^{(0)}_{\b A}\ket.
\ee
Inserting this result into Eq.~(\ref{nerf3f}) we obtain the 
nonequilibrium response function ($t>0$)
\bea
i\chi(t,0)&=&\sum_{\a\b\g}b^{\ast}_{\a}b_{\b}\left[
e^{i(\W_{\a\g}+\w_{P})t}\bra\Q^{(0)}_{\a A}|
\hat{d}|\Q^{(1)}_{\g B}\ket\right.
\nn\\
&+&\left.
e^{i(\W_{\a\g}-\w_{P})t}\bra\Q^{(1)}_{\a B}|
\hat{d}|\Q^{(0)}_{\g A}\ket\right]d_{\g\b}(0)-c.c.
\label{nerf3fsub}
\eea
For ultra-short probes $e(t)=e_{0}\d(t)$ the induced dipole moment 
$d_{p}(t)=e_{0}\chi(t,0)$, and we can easily deduce the position of 
the absorption regions in the
TR-PA spectrum from Eq.~(\ref{nerf3fsub}). 

\begin{figure}[tbp]
\centering
\includegraphics[width=\linewidth]{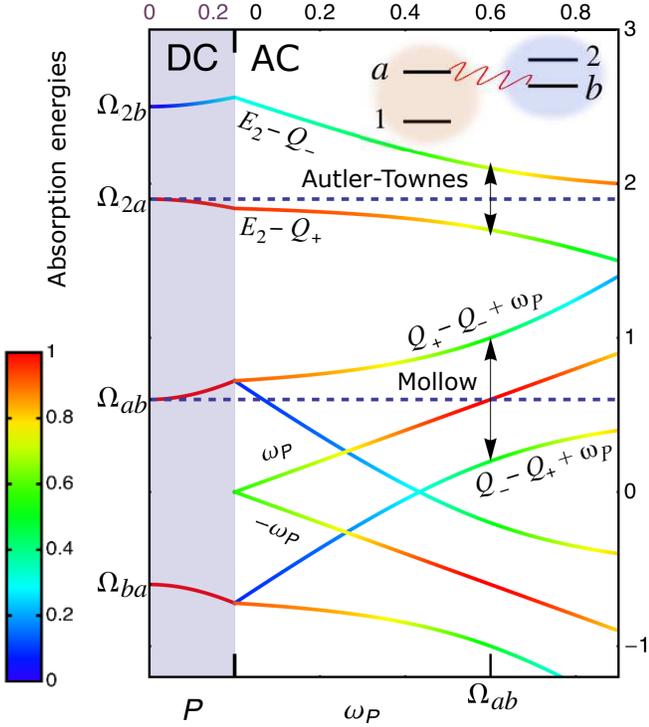}
\caption{(Color online) Absorption energies as a function of the pump 
intensity $P\in [0,0.2]$ for $\w_{P}=0$  [DC (grey) region] 
and as a function of $\w_{P}\in[0,0.9]$ 
for $P=0.2$ [AC region] of a four-level quantum system consisting of the states 
$1,a\in A$ with 
energy $E_1=0$, $E_a=21.2$, and states $b,2\in B$ with energies 
$E_b=20.6$ $E_2=23.1$ (energies in eV). The absorption at energy $\W_{12}$ is 
independent of $P$ and not shown. The color scale indicates 
the oscillator strength of the transition [$\bra\Q^{(0)}_{\a A}|
\hat{d}|\Q^{(1)}_{\g B}\ket$ for $\W_{\a\g}+\w_{P}$, $\bra\Q^{(1)}_{\a B}|
\hat{d}|\Q^{(0)}_{\g A}\ket$ for $\W_{\a\g}-\w_{P}$, and the sum of 
the two for $\w_{P}=0$, see Eq. (\ref{nerf3fsub})] normalized to 
$\mathfrak{D}\equiv\bra\Q_{\a}|\hat{d}|\Q_{\b}\ket$ with $\a=1,a$ and $\b=b,2$. In 
the AC region we also display the absorption energy $\w_{P}$ with 
oscillator strength $\sum_{\a}\bra\Q^{(0)}_{\a A}|
\hat{d}|\Q^{(1)}_{\g B}\ket$, and $-\w_{P}$ with oscillator strength $\sum_{\a}\bra\Q^{(1)}_{\a B}|
\hat{d}|\Q^{(0)}_{\g A}\ket$.}
      \label{illustrative-floquet}
\end{figure}

We conclude this Section by discussing the 
paradigmatic situation of a pump coupling only two states, say $a$ 
and $b$. 
Then $P_{\a\b}=P$ for $\a=a$ and $\b=b$, and zero 
otherwise. The quasi-energies are 
\be
Q_{\pm}=\frac{E_{a}+E_{b}+\w_{P}\pm
\sqrt{(E_{a}-E_{b}-\w_{P})^{2}+4P^{2}}}{2},
\label{quasienergy}
\ee
and for $\c\neq a,b$, 
$Q_{\c}=E_{\c}$ for $\c\in A$ and  $Q_{\c}=E_{\c}+\w_{P}$ for 
$\c\in B$. In the limit of zero pump intensity $P\to 0$ the 
quasi-energies $Q_{+}\to E_{a}$ and $Q_{-}\to E_{b}+\w_{P}$, as it 
should be. Let us analyze with the help of Fig.~\ref{illustrative-floquet} the various 
time-dependent contributions in Eq. (\ref{nerf3fsub}). For both $\a$ and 
$\g$ different  from $\pm$ the square bracket is nonvanishing 
only provided that $\a\in A$ [$B$] and $\g\in B$ [$A$]. More precisely, 
only the first [second] term is nonvanishing, and it contributes with an oscillating 
exponential of frequency 
$E_{\a}-E_{\g}-\w_{P}+\w_{P}=E_{\a}-E_{\g}$
[$E_{\a}+\w_{P}-E_{\g}-\w_{P}=E_{\a}-E_{\g}$]. Thus, 
the absorption energy between ``pump-invisible'' states is 
preserved (and hence not shown in Fig.~\ref{illustrative-floquet}).
The situation is more interesting for $\a\in A$ [$B$] a 
``pump-invisible'' state and $\g=\pm$ , see Fig.~\ref{illustrative-floquet} where 
the ``pump-invisible'' state is $\a=2\in B$. 
Again, only the first [second] term in the square bracket is 
nonvanishing and the corresponding oscillation frequency is 
$E_{\a}-Q_{\pm}+\w_{P}$
[$E_{\a}-Q_{\pm}$]. 
We observe that for $P=0$ the 
quasi-eigenstate $|\Q_{+}(t)\ket=|\Q_{a}\ket$ is in the $A$ subspace whereas the 
quasi-eigenstate $|\Q_{-}(t)\ket=e^{i\w_{P}t}|\Q_{b}\ket$
is in the $B$ subspace. 
Therefore the absorption at energy $E_{\a}-Q_{+}+\w_{P}$ 
[$E_{\a}-Q_{-}$ (see the line $E_{2}-Q_{-}$ in 
Fig.~\ref{illustrative-floquet})] is 
prohibited by the dipole selection rule. 
The pump field mixes $a$ and 
$b$, thereby giving rise to the appearance of a new peak for every 
equilibrium-forbidden transition between the ``pump-invisible'' state 
$\a\in B$ [$A$] and the state $b$ [$a$]. 
For $\w_{P}$ far from the resonant frequency 
$\W_{ab}\equiv E_{a}-E_{b}$
the (allowed) equilibrium transition 
$E_{\a}-E_{b}\ra E_{\a}-Q_{-}+\w_{P}$ [$E_{\a}-E_{a}\ra E_{\a}- 
Q_{+}$], thereby undergoing a shift 
known as the AC Stark shift.\cite{steck-book,bkkk.1969}
At the resonance frequency 
$\w_{P}=\W_{ab}$, 
the quasi-energies $Q_{\pm}=E_{a}\pm P$ and the equilibrium peak at 
$E_{\a}-E_{b}$ [$E_{\a}-E_{a}$] is replaced by two peaks of equal 
intensity at energy $E_{\a}-E_{b}\mp P$ [$E_{\a}-E_{a}\mp P$]. This spectral feature is known 
as the Autler-Townes doublet or splitting (since the original equilibrium peak 
appears split in two).\cite{at.1955,steck-book}
A similar analysis applies for $\a=\pm$ and $\g\in B$ [$A$]  a 
``pump-invisible'' state.
Finally we consider the 
contributions with $\a=+$ and $\g=-$  [$\a=-$ and $\g=+$] in 
Eq.~(\ref{nerf3fsub}). In this case both terms in the square brackets 
contribute and the equilibrium peak at energy $\W_{ab}=E_{a}-E_{b}$ 
[$\W_{ba}=E_{b}-E_{a}$] splits into two peaks at energy $Q_{+}-Q_{-}\pm\w_{P}$
[$Q_{-}-Q_{+}\pm\w_{P}$], see Fig.~\ref{illustrative-floquet}. It is worth noticing that this 
splitting and the Autler-Townes splitting have different origin. In 
the latter a prohibited transition becomes an allowed transition 
whereas in the former a genuinely new transition appear. At the 
resonance frequency the spectrum exhibits two peaks of equal 
intensity at energies $Q_{+}-Q_{-}\pm\w_{P}=2P\pm (E_{a}-E_{b})$ 
[$Q_{-}-Q_{+}\pm\w_{P}=-2P\pm (E_{a}-E_{b})$]. Therefore, in this 
case too the equilibrium peak at energy $\W_{ab}$ [$\W_{ba}$] appears 
split in two. Unlike in the Autler-Townes splitting, however, the 
distance between the peaks is $4P$ instead of $2P$, with the peak at 
energy $\W_{ab}+2P$ [$\W_{ba}-2P$] stemming from a  shift of the 
equilibrium peak at energy $\W_{ab}$ [$\W_{ba}$], and the peak 
at energy $\W_{ab}-2P$ [$\W_{ba}+2P$] stemming from the newly 
generated transition associated to the equilibrium peak at energy 
$\W_{ba}$ 
[$\W_{ab}$].
In addition to all the aforementioned absorption frequencies 
we have the pump frequency. In fact, for $\a=\g=\pm$ 
the square bracket in Eq.~(\ref{nerf3fsub}) is the sum of two 
oscillating exponentials with frequency $\pm\w_{P}$. Thus, 
at the resonance frequency the spectrum exhibits 
a three-prong fork structure known as the Mollow triplet:\cite{m.1969,steck-book} the side peaks at energy 
$\W_{ab}\pm 2P$ [$\W_{ba}\pm 2P$] and a peak in the middle at 
energy $\w_{P}=\W_{ab}$ [$-\w_{P}=\W_{ba}$].

\section{More analytic results and numerical simulations}
\label{numsec}

The analysis of the nonequilibrium response function $\chi$ carried out in 
the previous Section is useful for the physical interpretation of 
TR-PA spectra. In practice, however, it is numerically more 
advantageous to 
calculate the probe induced dipole moment $d_{p}$ directly. 
In this Section we present some more analytic results for 
systems consisting of a few levels and single out the effects on the 
TR-PA spectrum of a finite duration of the pump.

Let $\underline{\r}$ be the many-body density matrix in, e.g.,  the 
eigenbasis of 
the unperturbed Hamiltonian. The $(\a,\b)$ matrix element of 
$\underline{\r}$ is therefore $\bra\Q_{\a}|\hat{\r}|\Q_{\b}\ket$. In 
the same basis the matrix which represents the unperturbed 
Hamiltonian is diagonal and reads $\underline{H}={\rm diag}(\{E_{\a}\})$. We use the 
convention that an underlined quantity 
$\underline{O}$ represents the matrix 
of the operator $\hat{O}$ in the energy eigenbasis. The 
time-evolution of the density matrix is determined by the Liouville 
equation
\be
i\frac{d}{dt}\underline{\r}(t)=
[\underline{H}+\underline{H}_{P}(t)+\underline{H}_{p}(t),\underline{\r}(t)]-
\frac{i}{2}\{\underline{\G},\underline{\r}(t)\},
\label{liovilleeq}
\ee
where $\underline{\G}$ is a decay-rate matrix accounting 
for radiative, ionization and other decay-channels. In Eq.~(\ref{liovilleeq}) the 
matrices $\underline{H}_{P}$ and $\underline{H}_{p}$ represent the 
pump and probe interaction Hamiltonians, and the symbol ``$[\;,\;]$'' 
(``$\{\;,\;\}$'') is a (anti)commutator. 
We set the switch-on time of the pump at $t=0$ (hence the probe is 
switched on at time $t=\t$).
As we are interested 
in the solution of Eq.~(\ref{liovilleeq}) to lowest order in the 
probe field we write 
$\underline{\r}=\underline{\r}_{P}+\underline{\r}_{p}$, where 
$\underline{\r}_{P}(t)$ is the time-dependent density matrix with 
$\underline{H}_{p}=0$. Then, the probe-induced variation 
$\underline{\r}_{p}$ satisfies (omitting the time argument)
\be
i\frac{d}{dt}\underline{\r}_{p}=
[\underline{H}+\underline{H}_{P},\underline{\r}_{p}]
+[\underline{H}_{p},\underline{\r}_{P}]
-\frac{i}{2}\{\underline{\G},\underline{\r}_{p}\},
\label{dliovilleeq}
\ee
which should be solved with boundary condition 
$\underline{\r}_{p}(\t)=0$. For ultra-short probes 
$\underline{H}_{p}(t)=\d(t-\t)e_{0}\underline{d}$ and for times $t>\t$  
Eq.~(\ref{dliovilleeq}) simplifies to
\be
i\frac{d}{dt}\underline{\r}_{p}=
[\underline{H}+\underline{H}_{P},\underline{\r}_{p}]
-\frac{i}{2}\{\underline{\G},\underline{\r}_{p}\},
\label{dliovilleeq2}
\ee
which should be solved with boundary condition 
$\underline{\r}_{p}(\t)=-ie_{0}[\underline{d},\underline{\r}_{P}(\t)]$.
Once $\underline{\r}_{p}$ is known the probe-induced dipole moment can be 
calculated by tracing: 
$d_{p}=\Tr[\underline{\r}_{p}\underline{d}]$.

\subsection{Two-level system}

\begin{figure}[tbp]
\centering
\includegraphics[width=\linewidth]{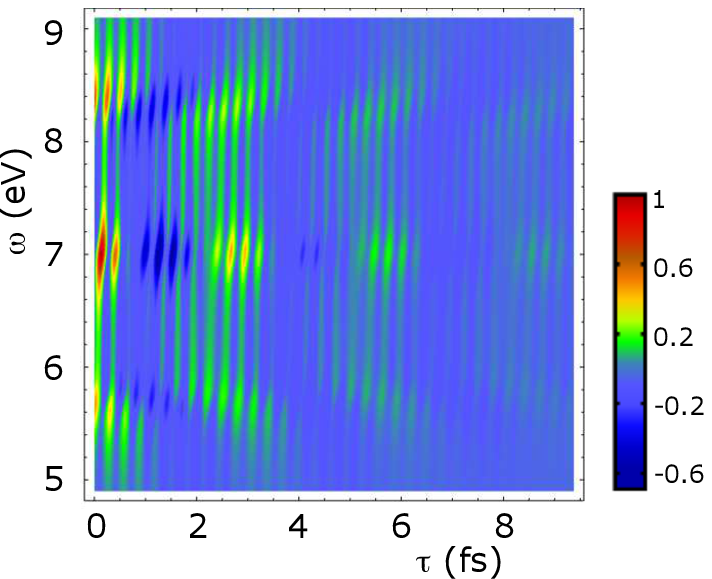}
\caption{(Color online) TR-PA spectrum (normalized to 
the maximum) for the
two level system described in the main text. The parameters are $\W_{ab}=E_{a}-E_{b}=7$, 
$P=0.7$, $\g_{a}=\g_{b}=\g=0.2$, $\g_{P}=0.02$ and $\w_{P}=\W_{ab}$. 
Energies in eV and times in fs. }
      \label{example1}
\end{figure}

We consider a two-level system with unpertubed
Hamiltonian $\underline{H}={\rm diag}(E_{a},E_{b})$, decay-rate 
matrix $\underline{\G}={\rm diag}(\g_{a},\g_{b})$,
and an exponentially decaying pump field which is suddenly switched-on at time 
$t=0$:
\be
\underline{H}_{P}(t)=\th(t)Pe^{-\g_{P}t}\left(\begin{array}{cc} 0 & e^{-i\w_{P}t} \\ 
e^{i\w_{P}t} & 0 \end{array}\right).
\ee
At time $t<0$ the state of the system is $|\Q_{a}\ket$ and hence the density matrix 
$\underline{\r}_{P}(t<0)={\rm diag}(1,0)$. In Fig.~\ref{example1} we show the 
TR-PA spectrum $\tilde{S}(\w)$ for $\g_{P}\ll \g$, 
$\w_{P}=\W_{ab}$ and for an ultra-short probe
$\underline{H}_{p}(t)=\d(t-\t)e_{0}\underline{d}$, where the dipole 
matrix has off-diagonal elements 
$d_{ab}=d_{ba}=\mathfrak{D}$ and zero on the diagonal.
 The spectrum is calculated from 
Eq.~(\ref{nedw}) without the inclusion of the quadratic term in 
$d_{p}$ (thin samples). We clearly distinguish a Mollow triplet for small $\t$.
As the delay increases the side peaks approach the peak at $\w_{P}$ 
and eventually merge with it. The $\t$-dependent shift of the side 
peaks is a consequence of the finite duration of the pump. 
In the resonant case we have found a simple analytic solution for the 
probe-induced dipole moment (for simplicity we consider $\g_{a}=\g_{b}=\g$)
\bea
d_{p}(t)\!\!&=&\!\!
2e_{0}\mathfrak{D}^{2}e^{-\g t}
\big[
\cos(\w_{P}\t)\sin(\w_{P}t)\cos\big(2P\frac{1-e^{-\g_{P}t}}{\g_{P}}\big)
\nn\\
\!\!&-&\!\!
\sin(\w_{P}\t)\cos\big(2P\frac{1-e^{-\g_{P}\t}}{\g_{P}}\big)
\cos(\w_{P}t)
\big],
\label{dp2levels}
\eea
for $t>\t$ and zero otherwise. 
As the Fourier transform $\tilde{d}_{p}(\w)$ is dominated by 
the behavior of $d_{p}(t)$ for times $\t\leq t\lesssim1/\g$ we 
study Eq.~(\ref{dp2levels}) in this range. 

We write $t=\t+s$ and define the function 
$\a(t)\equiv 2P(1-e^{-\g_{P}t})/\g_{P}$. 
For $\g_{P}\ll \g$ (this is 
the situation in Fig.~\ref{example1}) we can approximate
\be
\a(t)=
\a(\t)+2Pe^{-\g_{P}\t}s+\ldots
\ee
We then see that for a fixed $\t$ the first term in the square 
brackets in Eq.~(\ref{dp2levels}) oscillates  (as a function of $s$) at frequencies 
$\w_{P}\pm 2Pe^{-\g_{P}\t}$ whereas the second term oscillates at frequency 
$\w_{P}$. This explains the shrinkage of the splitting between the two 
side peaks with increasing $\t$ shown in Fig.~\ref{example1}. Another 
feature revealed by Eq. (\ref{dp2levels})
is that the side-peak intensity is proportional to 
$\cos(\w_{P}\t)$ while the central-peak intensity is  proportional to 
$\sin(\w_{P}\t)$ (antiphase). The Mollow triplet is therefore best 
visible only for delays $\t=(2n+1)\p/(4\w_{P})$, with $n$ integers.

\begin{figure}[tbp]
\centering
\includegraphics[width=\linewidth]{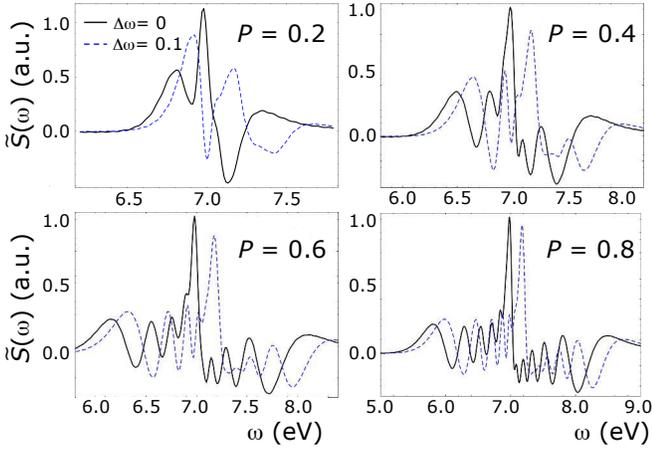}
\caption{(Color online) TR-PA spectrum (normalized to 
the maximum) for the two level system described in the main text 
at a delay $\t=\p/4\w_{P}$. The parameters are $\W_{ab}=E_{a}-E_{b}=7$, 
$\g_{a}=\g_{b}=\g_{P}=0.04$. Different panels refer to different pump 
intensity $P=0.2,\,0.4,\,0.6,\,0.8$ and the solid (dashed) line 
refers to the resonant (off-resonant) pump frequency 
$\w_{P}=\W_{ab}+\D\w$.
Energies in eV. }
\label{subsplitting}
\end{figure}

Equation~(\ref{dp2levels}) is valid for arbitrary $\g$ and $\g_{P}$. 
Hence, it can be used to investigate regimes other than $\g_{P}\ll \g$.  
In the opposite regime $\g_{P}\gg\g$ we can approximate 
$\a(t)\simeq 2P/\g_{P}$, and only one peak at 
frequency $\w_{P}$ is visible in the spectrum.
The intermediate regime, $\g_{P}\gtrsim \g$, is definitely the most 
interesting as it is characterized by a 
nontrivial  {\em sub-splitting} structure. In Fig.~\ref{subsplitting} we 
show the transient spectrum for $\g=\g_{P}=0.04$ at delay 
$\t=\p/(4\w_{P})$ for different pump strengths $P$. The results are 
obtained from the solution of Eq.~(\ref{dliovilleeq2}). We considered 
a resonant pump frequency, $\w_{P}=\W_{ab}$, as well as an off-resonant 
one, $\w_{P}=\W_{ab}+\D\w$. Although an analytic formula for $d_{p}$ 
exists in the non-resonant case too, it is much less transparent than 
Eq.~(\ref{dp2levels}) and not worth it to present. After a careful study of 
Eq.~(\ref{dp2levels}) we found that
the number of peaks in the frequency range $[-2P+\w_{P},2P+\w_{P}]$
grows roughly like $0.6\times P/\g_{P}$ and that the peak positions tend to 
accumulate around $\w_{P}$ (in the limit $ P/\g_{P}\ra\iif$ the 
frequency $\w_{P}$ becomes an accumulation point). The same 
qualitative behavior is observed for an off-resonant pump-frequency, 
the main difference being a shift by $\D\w$ of the sub-splitting 
structure. The sub-peaks are probably the most remarkable feature of 
the complicated functional dependence of the TR-PA spectrum on the pump and 
probe fields. The sub-peaks are
not related to transitions between light-dressed states and they
can be observed only for pump pulses of duration comparable with the 
dipole decay time.

\begin{figure}[tbp]
\centering
\includegraphics[width=\linewidth]{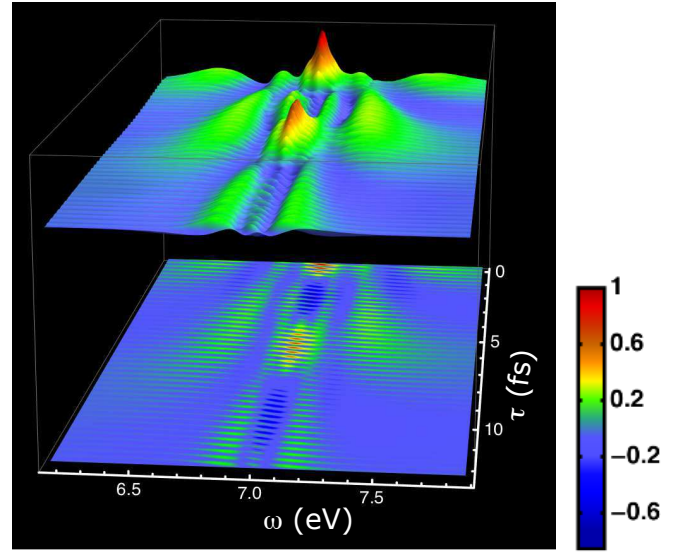}
\caption{(Color online) TR-PA spectrum (normalized to 
the maximum) for the two level system described in the main text.
The parameters are $\W_{ab}=E_{a}-E_{b}=7$, 
$\g_{a}=\g_{b}=\g_{P}=0.04$, $P=0.4$ and $\w_{P}=\W_{ab}$. 
Energies in eV and times in fs. }
\label{example2}
\end{figure}

The full transient spectrum in the intermediate regime is shown in 
Fig.~\ref{example2}. We use the same parameters as in 
Fig.~\ref{subsplitting} and set the pump intensity $P=0.4$. 
Thus, for $\t=\p/(4\w_{P})$ the spectrum is  identical to the one 
shown in the top-right panel of Fig.~\ref{subsplitting}. The 
sub-slitting structure evolves similarly as the main side-peaks. We 
observe periodic revivals of the sub-peaks whose positions  get progressively
closer to $\w_{P}$ and whose intensities decrease with $\t$. Another 
interesting feature is that for finite $\g_{P}$ the broadening of the peaks depends on 
$\t$. It is therefore important to take into account the finite 
duration of the pump when estimating the excitation life-times 
from the experimental widths.

\subsection{Three-level system}
\label{3levsec}

\begin{figure}[t]
\centering
\includegraphics[width=0.9\linewidth]{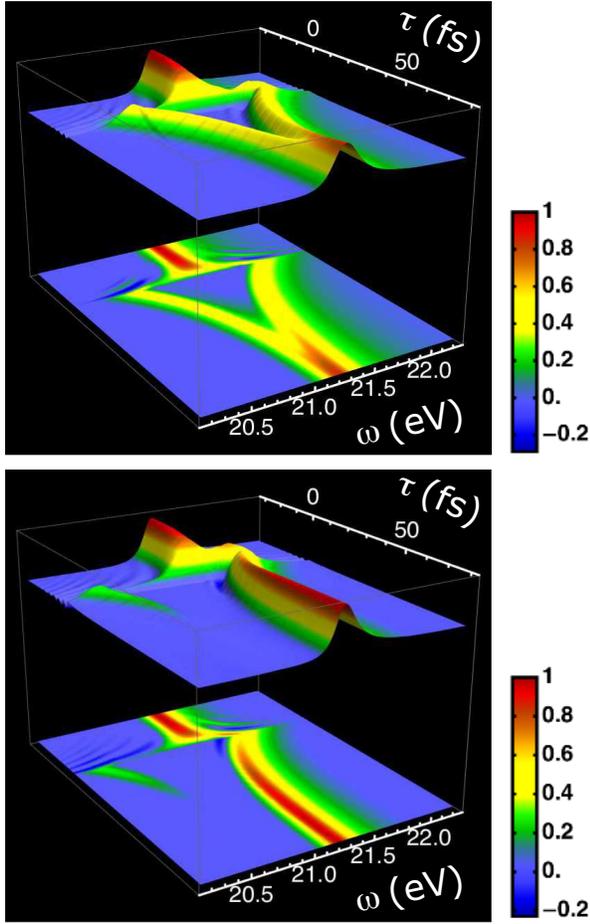}
\caption{(Color online) TR-PA spectrum (normalized to 
the maximum) for the three level system described in the main text. 
The parameters are $E_{1}=0$, $E_{b}=20.6$, $E_{a}=21.2$,  
$P=0.7$, $\g=0.2$, $\g_{P}=0.04$ and  
$\w_{P}=\W_{ab}$ (top panel) and $\w_{P}=\W_{ab}-0.5$ (bottom panel).
Energies in eV and time in fs.}
      \label{example3}
\end{figure}

We consider a three-level system with unperturbed Hamiltonian 
$\underline{H}={\rm diag}(E_{1},E_{b},E_{a})$ and decay-rate 
matrix $\underline{\G}={\rm diag}(0,\g,\g)$.  The system is perturbed by 
an exponentially decaying pump field which is suddenly switched-on at 
time $t=0$ and couples levels $a$ and $b$:
\be
\underline{H}_{P}(t)=\th(t)Pe^{-\g_{P}t}\left(\begin{array}{ccc} 
0 & 0 & 0 \\ 
0 & 0 & e^{i\w_{P}t} \\ 
0 & e^{-i\w_{P}t} & 0 \end{array}\right).
\ee
In the absence of the probe 
the system is in the ground state $|\Q_{1}\ket$ before the pump is 
switched on, hence $\underline{\r}_{P}(t<0)={\rm diag}(1,0,0)$. At 
time $\t$ we switch on an ultrashort probe field 
$\underline{H}_{p}(t)=e(t)\underline{d}$ where $e(t)=e_{0}\d(t-\t)$ 
and, for simplicity, we take the dipole matrix of the form
\be
\underline{d}=
\left(\begin{array}{ccc} 0 & 0 & \mathfrak{D} \\ 0 & 0 & 0 \\ 
\mathfrak{D} & 0 & 
0 \end{array}\right)
\label{dipmat3level}
\ee
(we therefore neglect the matrix element 
$\bra\Q_{a}|\hat{d}|\Q_{b}\ket$). 
For $\w_{P}\simeq E_{a}-E_{b}$ in the 
near-infrared region and $E_{a}-E_{1}$ in the extreme ultraviolet 
a similar model has been considered by several authors in studies 
of TR-PA of He atoms (in this case $1=1s^{2}$, $b=1s2s$, $a=1s2p$ represent the first 
three levels of 
He),\cite{gbts.2011,rthzetal.2011,cl.2012,pl.2012,cbbmetal.2012,pbbmnl.2013,cwgs.2013,wcgs.2013,cwcc.2014}
although the time-dependent probe and 
pump fields were different.
  
In Fig.~\ref{example3} we show the 
TR-PA spectrum for a resonant (top panel) and 
off-resonant (bottom panel) pump frequency $\w_{P}$. 
The spectrum is again calculated from Eq.~(\ref{nedw}) without the inclusion of the quadratic 
term in $d_{p}$ (thin samples).
For negative $\t$ 
we only see the equilibrium peak at frequency $\W_{a1}$, 
which corresponds to the excitation from the occupied level $1$ to the 
empty level $a$ (given the choice of the dipole matrix in 
Eq.~(\ref{dipmat3level}) this is the only possible transition). For 
small positive $\t$ we recognize the Autler-Townes splitting 
discussed in Section \ref{monoprobesec}. The finite duration of the 
pump causes the collapse of the Autler-Townes splitting as $\t$ 
increases; for $\t\ra\iif$ we eventually recover the equilibrium 
PA spectrum. The $\t$-dependent shift of the Autler-Townes peaks can 
be study quantitatively in the resonant case. In fact, the 
differential equation for $\underline{\r}_{p}$,  see
Eq.~(\ref{dliovilleeq2}), can be solved analytically 
for $\w_{P}=\W_{ab}$ and the corresponding probe-induced 
dipole moment reads
\bea
d_{p}(t)&=&\th(t-\t)
2e_{0}\mathfrak{D}^{2}e^{-\g(t-\t)}
\sin\big(\W_{1a}(t-\t)\big)
\nn\\&\times&
\cos\big( Pe^{-\g_{P}\t}
\frac{1-e^{-\g_{P}(t-\t)}}{\g_{P}}\big)
\label{atanalytic}
\eea
Interestingly $d_{p}(\t+s)$ depends on $\t$ only through the 
exponentially renormalized pump intensity $Pe^{-\g_{P}\t}$: 
the PA spectrum at finite $\t$ and pump intensity $P$  is the same as the PA spectrum 
at $\t=0$ and pump intensity $Pe^{-\g_{P}\t}$. Therefore, the 
Autler-Townes splitting follows the exponential decay of the 
pump, in agreement with the numerical simulation in 
Fig.~\ref{example3}.

In the TR-PA spectrum of Fig.~\ref{example3} we have $\g_{P}\ll\g$. However, the 
analytic solution in Eq.~(\ref{atanalytic}) is valid for all $\g$ and 
$\g_{P}$. Like in the two-level system a sub-splitting structure 
emerges in the intermediate regime 
$\g\simeq \g_{P}$ (not shown). A similar finding was recently found for 
trigonometric and square pump-envelops.\cite{wcgs.2013}

The TR-PA spectra shown so far have been calculated under the assumption 
that the total electric probe field acting on the electrons is the same as 
the external (bare) field. As discussed in Section \ref{monoprobesec}, this 
approximation makes sense for thin samples. For samples of thickness 
much larger than the inverse transition energies of 
interest, the Liouville equation 
for the density matrix, Eq. (\ref{liovilleeq}), should be coupled to 
the equation for the total electric field, Eq.~(\ref{poleq2}).\cite{mukamel-book,zag.1995,bsy.2007,gbts.2011,cl2.2012,csg.2012,cl.2013,wcsg.2013,pbbmnl.2013,oyi.2009,yssob.2012,wlbsetal.2014,ds.2014} 
To appreciate the qualitative difference introduced by a 
self-consistent treatment of the probe field we consider again the system of 
Fig.~\ref{example3} and add to the bare  
$\d$-like probe an induced exponentially decaying planewave of 
frequency $\W_{a1}$. We enforce (for simplicity) monochromaticity on the probe 
Hamiltonian (or equivalently we work in the rotating wave 
approximation) and take a total probe field 
$\bcallE_{p}(t)=(\callE_{x}(t-\t),\callE_{y}(t-\t),0)$, where 
the components 
\bea
\callE_{x}(t)&=&e_{0}\d(t)+\th(t)\l e^{-\g_{P}t}\cos\W_{a1}t
\nn\\
\callE_{y}(t)&=&\th(t)\l e^{-\g_{P}t}\sin\W_{a1}t\eea
decay on the same 
time scale of the pump field. Choosing the dipole components 
$\underline{\bld}=(\underline{d}_{x},\underline{d}_{y},\underline{d}_{z})$ with 
\be
\underline{d}_{x}=
\left(\begin{array}{ccc} 0 & 0 & \mathfrak{D} \\ 0 & 0 & 0 \\ 
\mathfrak{D} & 0 & 
0 \end{array}\right)\quad;\quad
\underline{d}_{y}=
\left(\begin{array}{ccc} 0 & 0 & i\mathfrak{D} \\ 0 & 0 & 0 \\ 
-i\mathfrak{D} & 0 & 
0 \end{array}\right),
\label{dipmat3levelxy}
\ee
the probe Hamiltonian reads
\be
\underline{H}_{p}(t)=\bcallE(t)\cdot\underline{\bld}
=\mathfrak{D}\left(\begin{array}{ccc} 
0 & 0 & \callE_{p}(t-\t) \\ 
0 & 0 & 0 \\ 
\callE^{\ast}_{p}(t-\t) &  & 0 \end{array}\right),
\label{scprobe}
\ee
with
$\callE_{p}(t)=e_{0}\d(t)+\th(t)\l e^{i\W_{a1}t-\g_{P}t}$.
Our modelling of the induced probe field is 
based on the self-consistent results of Ref.~\onlinecite{pbbmnl.2013}. 
We instead assume that the pump field does not need to be dressed. 

\begin{figure}[t]
\centering
\includegraphics[width=0.9\linewidth]{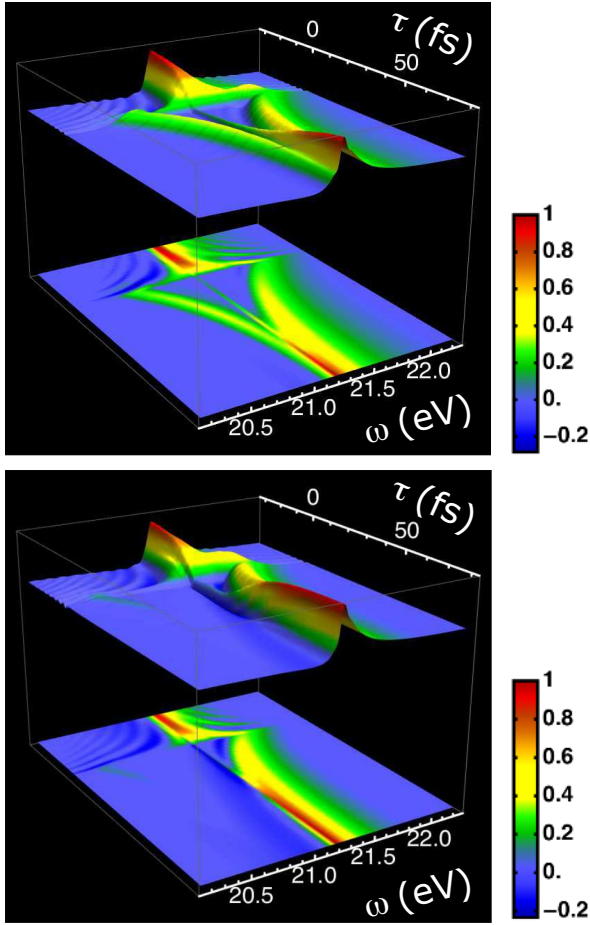}
\caption{(Color online) Linear term in $d_{p}$ of the TR-PA spectrum (normalized to 
the maximum) of Eq.~(\ref{nedw}) for the three level system described in the main text. 
The probe Hamiltonian is given in Eq.~(\ref{scprobe}) with 
$\l/e_{0}=0.01$.
The rest of the parameters are the same as in Fig.~\ref{example3}. 
The pump frequency is $\w_{P}=\W_{ab}$ (top panel) and 
$\w_{P}=\W_{ab}-0.5$ (bottom panel).
Energies in eV and time in fs.}
      \label{example4}
\end{figure}
The formula for the spectrum, 
Eq.~(\ref{nedw}), has beed derived for linearly polarized probe 
fields. It is straightforward to show that for 
$\ble=\sum_{n}\bgve_{p}^{(n)}e^{(n)}$ the generalization is
\bea
\tilde{S}(\w)=-2\sum_{n}\Im\left(\!\w\,\tilde{e}^{(n)\ast}(\w)\tilde{d}^{(n)}_{p}(\w)\!\right)
-\frac{2\p}{\callS c}\left|\w\,\tilde{\bld}_{p}(\w)\right|^{2}\!\!.
\label{nedwmp}
\eea
where $\tilde{d}^{(n)}_{p}\equiv \bgve_{p}^{(n)}\cdot 
\tilde{d}^{(n)}_{p}$. In Fig.~\ref{example4} 
we display the first term (linear in $d_{p}$) of the transient PA 
spectrum of Eq.~(\ref{nedwmp}). The main difference with the spectrum 
in Fig.~\ref{example3} is the appearance of an extra peak at 
frequency $\W_{1a}$. It is therefore the induced probe field 
to generate the central peak. Furthermore, the height of the central peak 
increases monotonically (no coherent oscillations), in 
agreement with the results of Section~\ref{monoprobesec}. One more remark is about the exponential 
shrinkage of the Autler-Townes splitting already observed in 
Fig.~\ref{example3}. According to our calculations, the Autler-Townes splitting and the pump 
intensity decay on the same time-scale, see Eq.~(\ref{atanalytic}). 
This implies that a spectral shift can occur only provided that 
the amplitude of the pump field is delay dependent as is, for 
instance, the case in thick samples where the pump is dressed 
by an exponentially decaying dipole moment.   

\begin{figure}[t]
\centering
\includegraphics[width=0.9\linewidth]{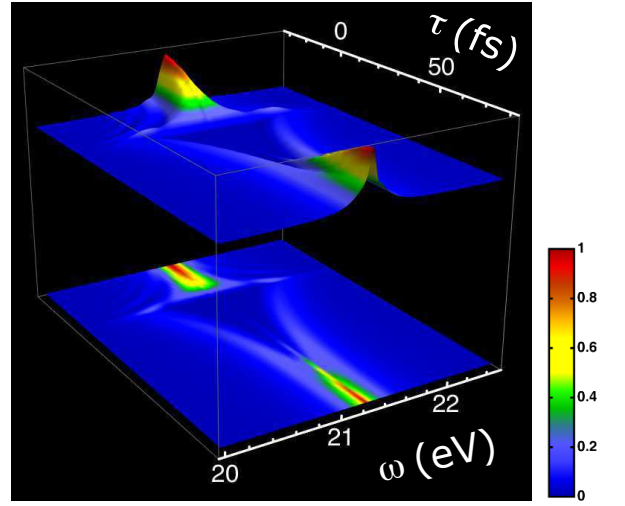}
\caption{(Color online) Plot of $|\w\,\bar{d}_{p}(\w)|^{2}$ (normalized to 
the maximum) for the three level system described in the main text. 
Same parameters as in the top panel of Fig.~\ref{example4}.
Energies in eV and time in fs.}
      \label{figd2}
\end{figure}

For thick samples the quadratic term in $d_{p}$ in the TR-PA
spectrum of Eq.~(\ref{nedwmp}) can become relevant. This term is always 
{\em negative} and hence it can either suppress a positive peak or 
even turn a positive peak into a negative one. The induced dipole 
moment scales linearly with the sample volume $V=\callS L$. If we 
introduce the dipole density per unit volume $\bar{d}_{p}=d_{p}/V$, 
then the TR-PA spectrum in Eq.~(\ref{nedwmp}) can be rewritten as
\bea
\frac{\tilde{S}(\w)}{V}=-2\sum_{n}\Im\left(\!\w\,\tilde{e}^{(n)\ast}(\w)\tilde{\bar{d}}^{(n)}_{p}(\w)\!\right)
-\frac{2\p 
L}{c}\left|\w\,\tilde{\bar{\bld}}_{p}(\w)\right|^{2}\!\!,\;\;
\label{nedwdipdens}
\eea
from which we see that the contribution of the last term grows linearly 
with the sample thickness. The quantity 
$|\w\,\tilde{\bar{\bld}}_{p}(\w)|^{2}$ is shown in 
Fig.~\ref{figd2} in the resonant case $\w_{P}=\W_{ab}$. As expected, 
the spectral regions where the linear (top panel of 
Fig.~\ref{example4}) and quadratic (Fig.~\ref{figd2}) terms are 
nonvanishing are the same. Nevertheless, the mathematical 
structure of the peaks is very different, as it should be. In fact,  
in the 
absence of damping $\tilde{d}_{p}(\w)$ is the sum of Dirac 
$\d$-functions and hence its square is the sum of Dirac 
$\d$-functions squared. 

\section{Summary and Conclusions}
\label{summsec}

We have provided a detailed analysis of the nonequilibrium 
dipole response function $\chi$, the fundamental physical 
quantity to be calculated/simulated for interpreting  
TR-PA spectra. 
Exact and general properties of $\chi$ have been elucidated and 
then related to transient spectral 
features.
In the nonoverlapping regime the 
height of the absorption peaks are strongly affected by the shape of 
the probe pulse. For ultrashort probes the peak heights exhibit quantum beats 
as a function of the delay, a signature of the coherent electron 
motion  in the 
nonstationary state created by the pump. As the probe duration 
increases the effects of coherence are progressively washed out, and 
the spectrum is progressively 
suppressed away from the probe frequency. The absorption regions are instead independent of the 
delay and occur in correspondence of the neutral excitation energies (not necessarily 
involving the ground-state) of 
the equilibrium system.
For overlapping pump and probe the absorption regions
cease to be an intrinsic property of the equilibrium system
and, more generally, the interpretation of the transient spectrum becomes  
intricate. Analytic results for everlasting periodic pump fields are 
available and at the same time useful for the interpretation of 
TR-PA spectra. The Lehmann-like representation of $\chi$ in terms of 
light-dressed states provide a unifying framework for a  variety of 
well known phenomena, e.g., 
the AC-Stark shift, the Autler-Townes splitting, the 
Mollow triplet, the photon replicas, etc. 

The effects of the finite duration of the pump pulse are difficult to 
address in general terms. We have  considered 
the two- and three-level systems extensively studied in the literature
and derived an exact analytic expression for the time-dependent  
probe-induced dipole moment. Our 
solution shows that for strong enough pump intensities a rich 
sub-splitting structure emerges, in agreement with the recent 
theoretical findings  in Ref.~\onlinecite{wcgs.2013}.  
We also find agreement with recent experimental results on He: for 
long (induced) probe fields, like those occurring in thick samples, 
the absorption peak at the probe frequency does not exhibit 
coherent oscillations.\cite{pbbmnl.2013}


\begin{thebibliography}{112}
\bibitem[{not({\natexlab{a}})}]{note2}
\bibinfo{note}{For thick samples the total field $\callE$ is not spatially
  uniform and Eq.~(\ref{scalarkubo}) involves an integral over space too.}
\bibitem[{not({\natexlab{b}})}]{note1}
\bibinfo{note}{The same result follows from the direct solution of the
  Schr\"odinger equation
  $i\frac{d}{dt}|\Q(t)\ket=\big(\hat{H}+\hat{H}_{P}(t)\big)|\Q(t)\ket$.
  Projecting this equation onto subspaces $A$ and $B$ one finds
  $i\frac{d}{dt}|\Q_{A}(t)\ket=\hat{H}_{A}|\Q_{A}(t)\ket+\hat{P}e^{-i\w_{P}t}|\Q_{B}(t)\ket$
  and
  $i\frac{d}{dt}|\Q_{B}(t)\ket=\hat{H}_{B}|\Q_{B}(t)\ket+\hat{P}^{\dag}e^{i\w_{P}t}|\Q_{A}(t)\ket$.
  Writing $|\Q_{B}(t)\ket=e^{i\w_{P}t}|\F_{B}(t)\ket$ the problem is mapped
  onto a Schr\"odinger equation for time-independent Hamiltonians:
  $i\frac{d}{dt} \left( \begin{array}{c} |\Q_{A}(t)\ket \\ |\F_{B}(t)\ket
  \end{array}\right) =\left( \begin{array}{cc} \hat{H}_{A} & \hat{P} \\
  \hat{P}^{\dagger} & \hat{H}_{B} + \w_{P} \end{array} \right) \left(
  \begin{array}{c} |\Q_{A}(t)\ket \\ |\F_{B}(t)\ket \end{array}\right)$.}
\end{thebibliography}

\end{document}